\documentclass[preprint,review,3p]{elsarticle}

\usepackage{graphics}
\usepackage{graphicx}
\usepackage{subfigure}
\usepackage[numbers]{natbib}

\usepackage{psfrag}
\usepackage{cancel}
\usepackage{amssymb}
\usepackage{amsmath}
\usepackage{mathrsfs}

\usepackage{amsthm}
\usepackage{xcolor}
\usepackage{rotating}
\usepackage{lscape}
\usepackage{pdflscape}
\usepackage{booktabs}
\usepackage{multirow}

\usepackage{lineno}
\usepackage{float}
\usepackage{epstopdf}
\usepackage{changes}
\usepackage{lipsum}

\colorlet{Changes@Color}{red}

\usepackage{hyperref}

\usepackage{ragged2e}

\begin{document}

\begin{frontmatter}




\title{Analytical solutions for the acoustic field in thin annular combustion chambers with linear gradients of cross-sectional area and mean temperature}
%
\author[af1]{Dongbin~Wang}
\ead{bh$\_$wangdb@buaa.edu.cn}
\author[af1]{Jiaqi~Nan}
\ead{nan$\_$jiaqi@buaa.edu.cn}
\author[af1,af2]{Lijun~Yang}
\ead{yanglijun@buaa.edu.cn}
\author[af3]{Aimee S.~Morgans}
\ead{a.morgans@imperial.ac.uk}
\author[af1,af2]{Jingxuan~Li\corref{cor4}}
\ead{jingxuanli@buaa.edu.cn}
\address[af1]{School of Astronautics, Beihang University, Beijing 100191, China.}
\address[af2]{Beijing Advanced Innovation Center for Big Data-based Precision Medicine, Beihang University, Beijing, 100083, China.}
\address[af3]{Department of Mechanical Engineering, Imperial College London, London, SW7 2AZ, UK.}
\cortext[cor4]{Corresponding author at: School of Astronautics, Beihang University, Beijing 100191, China.}

\begin{abstract}

Predictions of thermoacoustic instabilities in annular combustors are essential but difficult. Axial variations of flow and thermal parameters increase the cost of numerical simulations and restrict the application of analytical solutions. This work aims to find approximate analytical solutions for the acoustic field in annular ducts with linear gradients of axially non-uniform cross-sectional area and mean temperature. These solutions can be applied in low-order acoustic network models and enhance the ability of analytical methods to solve thermoacoustic instability problems in real annular chambers.
A modified WKB method is used to solve the wave equation for the acoustic field, and an analytical solution with a wide range of applications is derived. 
The derivation of the equations requires assumptions such as low Mach number, high frequency and small non-uniformity. Cases with arbitrary distributions of  cross-sectional surface  area and mean temperature can be solved by the piecewise method as long as the assumptions are satisfied along the entire chamber.

\end{abstract}
\end{frontmatter}
%
\section{Introduction}
\label{sec:1} 
Thermoacoustic instability is of concern in applications such as gas turbines and aero-engines, leading to research into instability problems in combustion chambers \citep{Ruan_AST_2020,Chen_AST_2019} and the acoustic field in combustors \citep{Lilei_AST_2015,LiLei_AST_2018,Si_AST_2013,Kierkegaard_AST_2010}. Experiments and detailed numerical simulations are typically very expensive \citep{Cheng_AST_2021}, making them less suitable for industrial design and prediction. Low-order acoustic network models are widely used in the prediction and model-based control of thermoacoustic instabilities. They treat a complex combustor geometry as a set of connected ducts, each with simple or uniform geometry and flow field \citep{Dowling_JPP_2003,Dowling_ARFM_2005,Han_CNF_2015,Li_CNF_2017}. This greatly reduces the computational cost compared to methods which directly simulate the coupling effect of the unsteady heat release rate and the acoustic field by using full 3D compressible computational fluid dynamics  \citep{Poinsot_PCI_2017,Selle_CNF_2004}. When dealing with a geometry whose cross-sectional area and flow field change gradually, the chamber can be segmented into a series of elements whose individual axial lengths are sufficiently smaller than the dominant acoustic wavelengths.  The cross-sectional area and flow field in each element is then assumed uniform, so that analytical solutions for the acoustic field within each element can be obtained directly and used \citep{Li_OSCILOS,yang2019systematic,Yang_AST_2021}. However, if it is possible to directly derive analytical solutions for the acoustic fields in these complex geometries, the computing cost could be further reduced and it would also be possible to obtain more physical insights into the mechanisms.

For ducts with continuously changing cross-sectional area, it is possible to derive exact analytical solutions for the acoustic field when the axial variation of the geometric profile takes some specific forms,   e.g., linear functions or exponential functions \citep{Webster_PNAS_1919,Salmon_JASA_1946}. 
In order to derive exact analytical solutions for the acoustic field in ducts with temperature gradients and no mean flow, Sujith and his colleagues \citep{Sujith_JSV_1995, Kumar_JSV_1998, Kumar_JASA_1997} derived the acoustic wave equation based on the linearised Euler equation of the unsteady flow, which was then converted into an ODE with exact analytical solutions using a variable transformation method.
Karthik \citep{Karthik_JASA_2000} and Veeraragavan \citep{Veeraragavan_AIAA_2006} conducted similar derivations for low Mach number conditions. All the wave equations have exact analytical solutions in the form of Bessel functions or hypergeometric functions. This method provides a clever way to deal with non-uniform ducts, but the derivation depends on the axial variation of the mean temperature gradient  and has limited applicability. The acoustic field in cylindrical and annular ducts with varying mean temperature and non-mean flow is solved in Ref. \citep{Li_AST_2020} by separation of variables.

When the geometry becomes more complex and the above method is unsuitable, it is possible to segment the geometry into a series of elements and fit the geometry of each  using a low-order polynomial function \citep{Pagneux_JASA_1996}. For instance, Pillai  proposed an analytical method supported by Fuch's theorem and Frobenius series to determine the low-frequency acoustic  field for arbitrary shape and area \citep{Pillai_JASA_2019}. In this method, the  length of each element should be much smaller than the radius of curvature of each element's outer wall.

The WKB (Wentzel-Kramers-Brillouin) method is an approximate method to derive the analytical solutions of differential equations \citep{Henrick_JASA_1983}. 
Cummings \citep{Cummings_JSV_1977} proposed a modified WKB approximation and derived analytical solutions for the acoustic field in a duct with a rectangular section and axial temperature gradient. This method was then improved by Li \& Morgans \citep{Li_JSV_2017} and applied to ducts with more complex geometries and mean temperature distributions. Yeddula \& Morgans \citep{Yeddula_JSV_2020} derived solutions for the acoustic field in one-dimensional ducts with both arbitrarily varying cross-sectional area and temperature gradient. Further studies investigated the effects of mean flow and mean temperature gradient on acoustic absorption and generation within the duct \citep{Yeddula_JSV_2021}. Rani \& Rani \citep{Rani_JSV_2018} derived two approximate WKB solutions for quasi-one-dimensional ducts with arbitrary mean properties and mean flow, while the nonlinear temporal dynamics of acoustic oscillations was discussed by Basu \citep{Basu_JSV_2022}. The acoustic field in a three-dimensional rectangular duct with a temperature gradient and mean flow has also been derived \citep{Nan_AST_2021}. All the WKB approximate solutions agree well with numerical results under certain assumptions and application to chambers with more complex non-uniform profiles is anticipated.

The present work extends our previous research by applying the modified WKB approach to a thin annular combustion chamber. 
Annular combustion chambers are typically the preferred geometry in aero-engines. The radial gap of the annular chamber is typically much smaller than the axial length and circumference, leading to the annular chamber being considered ``thin'' such that radial acoustic modes are neglected. As the  cross-sectional area and mean flow field within the thin annular chamber are typically not axially uniform, a simple method is to segment the complex geometry into a series of sections with simple geometries and flow fields. In this work, we assume that each section contains a linearly varying cross-sectional area and temperature gradient. Based on this assumption, the derivation of the analytical solution for the acoustic field is obtained.


%
%
%

This paper is organised as follows. The derivation of the relevant wave equation and its subsequent solution using the modified WKB approximation are described in section~\ref{sec:2}. In section~\ref{sec:3}, the configuration of the validation geometry and operating conditions are presented. Section~\ref{sec:4} compares the analytical and numerical solutions. The acoustic field for a case with more complex area and temperature profiles is then calculated by applying a linear piecewise approach. Conclusions are drawn in section~\ref{sec:5}.


\section{Acoustic wave equation and analytical solutions}
\label{sec:2} 

\subsection{The wave equation for the acoustic field in ducts with axially linear variations}
\label{subsec:2:1}
\begin{figure}[ht]
\centering
	\includegraphics[height=5cm]{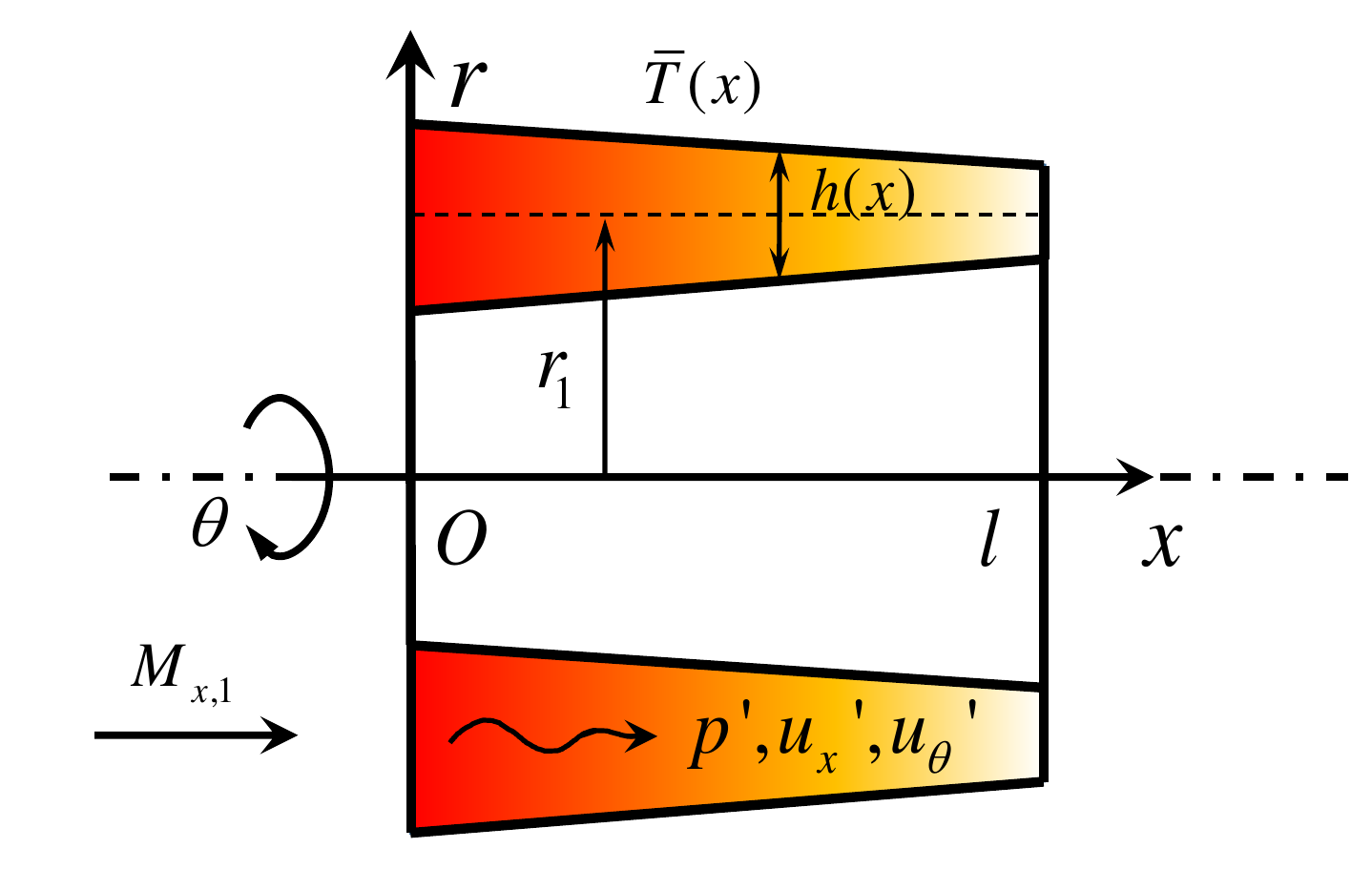}
	\caption{Sketch of a thin annular combustion chamber with a cross-sectional area and mean temperature gradient which both vary linearly in the axial direction.} 
	\label{fig:sketch1}
\end{figure}

In cylindrical polar coordinates (Fig.~\ref{fig:sketch1}), a thin annular combustion chamber with an inviscid perfect gas is considered. 
The narrow-gap assumption is applied and a constant mean radius is assumed, indicating that $h \ll r$ and $r \equiv r_1$. The length of the duct is $l$ and the Mach number at the inlet is $M_{x,1}$. The gap height $h$ changes linearly along the axial coordinate $x$ and a linear profile of mean temperature is also considered. Two parameters $\alpha_{h}$ and $\alpha_{\overline{T}}$ are used to describe the variation of the gap height and the mean temperature,
\begin{equation}
\label{eq:alpha_h_Ta}
\alpha_{h} = \arctan\left(\frac{h_{2} - h_{1}}{2 l} \right), \quad
\alpha_{\overline{T}} = \frac{\overline{T}_{2} - \overline{T}_{1}}{l}.
\end{equation}
The subscripts $`1'$ and $`2'$ represent parameters at the duct inlet and outlet, respectively. The profiles of the gap height $h$ and mean temperature  $\overline{T}$ can be expressed as,
\begin{equation}
\label{eq:h_Ta}
h(x) = h_{1} + 2 x \cdot \tan \alpha_{h} , \quad
\overline{T}(x) = \overline{T}_{1} + x \cdot \alpha_{\overline{T}}.
\end{equation}
The acoustic field in ducts with more complex geometries and temperature distributions can be solved by applying a piecewise linear function (PLF) approach and using the solutions for linear profiles, as in our previous work \citep{Li_JSV_2017}.

Equations for the conservation of mass, axial-momentum, circumferential-momentum and energy as well as the equation of state in the cylindrical polar coordinates can be expressed as follows,
\begin{equation}
\label{eq:mass_thin}
A\frac{\partial \rho}{\partial t}   + \frac{\partial \left(A\rho u_x^{}\right)}{\partial x}
+   \frac{1}{r}\frac{\partial \left(\rho A u_\theta^{}\right)}{\partial \theta}   = 0,
\end{equation}
\begin{equation}
\label{eq:momentum_x_thin}
\frac{\partial u_x^{}}{\partial t} +  u_x^{} \frac{\partial u_x^{}}{\partial x} +  \frac{u_\theta^{}}{r} \frac{\partial u_x^{}}{\partial \theta} + \frac{1}{\rho}\frac{\partial p }{\partial x}  = 0 ,
\end{equation}
\begin{equation}
\label{eq:momentum_theta_thin}
\frac{\partial u_\theta^{}}{\partial t} +   u_x^{} \frac{\partial u_\theta^{}}{\partial x} +  \frac{u_\theta^{}}{r} \frac{\partial u_\theta^{}}{\partial \theta} + \frac{1}{r\rho}\frac{\partial p }{\partial \theta}  = 0 ,
\end{equation}
\begin{equation}
\label{eq:energy_thin}
\frac{A}{\gamma p}\left(\frac{\partial p}{\partial t}       +   u_x^{} \frac{\partial p}{\partial x}
+\frac{u_\theta^{}}{r} \frac{\partial p}{\partial \theta}\right)
+  \frac{\partial \left(A u_x^{}\right)}{\partial x} 
+ \frac{1}{r}\frac{\partial \left( A u_\theta^{} \right)}{\partial \theta}  = \frac{\gamma- 1}{\gamma p} A\dot{q} ,
\end{equation}
\begin{equation}
\label{eq:ideal_gas}
p = \rho R_g T ,
\end{equation}
where $p, \rho$ and $T$ are pressure, density and temperature, respectively. $A = 2 \pi r_1 h$ is the cross-sectional area of the annular gap. $u_{x}$ and $u_{\theta}$ are velocities in the longitudinal and circumferential directions. $R_g$ and $\gamma$ represent the gas constant and the ratio of specific heats. $\dot{q}$ denotes the heat flux. Following our previous research, only the mean value of $\dot{q}$ is accounted for, leading to a distribution of mean temperature; the perturbation in $\dot{q}$ is neglected. 

%
The linearised Euler equations (LEEs) can be derived from  Eqs.~\eqref{eq:mass_thin}-\eqref{eq:energy_thin} based on linearisation and separation of variables. Each flow parameter is considered as the superposition of a time-averaged part marked with $\bar{(~)}$ and a fluctuating part marked with $(~)^\prime$, e.g., $p=\bar{p} + p^\prime$. 
A harmonic perturbation is considered, such that the perturbation can be expressed as,  e.g., $p' = \hat{p} (x) \mathrm{e} ^{\mathrm{i}\omega t + \mathrm{i} n \theta}$, where $\mathrm{i} = \sqrt{-1}$, $\omega$ and $n$ are the angular frequency and circumferential wavenumber. 
 By assuming that the radial and circumferential mean flow can be neglected (both being much smaller than the axial mean flow $\bar{u}_{x}$), the LEEs with four equations follow directly from Eqs.~\eqref{eq:mass_thin}-\eqref{eq:energy_thin} (see \ref{sec:Appendix_LEEs_0_4}).

A low Mach number assumption is made such that the entropy and vorticity waves can be neglected \citep{Dowling_JPP_2003}. The entropy fluctuations generated by the interaction of acoustic waves with the non-uniform cross-sectional area and mean temperature in the axial-momentum equation have been shown to have an insignificant effect on the acoustic field \citep{Li_JSV_2017,Li_JSV_2020}. The entropy term can be expressed as,
\begin{equation}
\label{eq:entropy}
\hat{\sigma} = \frac{\hat{p}}{ \gamma \bar{p}} - \frac{\hat{\rho}}{\bar{\rho}}.
\end{equation}

The time-averaged parts of Eqs.~\eqref{eq:mass_thin}-\eqref{eq:ideal_gas} can be obtained by eliminating spatial derivatives of the mean parameters, this proving useful across the entire derivation (see \ref{sec:Appendix_D_param}). Substituting the spatial derivative terms, the equations for the pressure perturbation $\hat{p}$, axial velocity perturbation $\hat{u}_x$ and circumferential velocity perturbation $\hat{u}_\theta$ are as follows, 
\begin{equation}
\label{eq:energy_LEEs}
\bar{u}_x^{} \frac{\mathrm{d} \hat{p}}{\mathrm{d} x}  +  
\left(\mathrm{i} \omega  +  \frac{\gamma \bar{u}_{x}}{1-\gamma M_{x}^{2}} \left(\beta - \gamma M_{x}^{2} \alpha \right) \right) \hat{p}  
+  \gamma \bar{p} \frac{\mathrm{d} \hat{u}_x{}}{\mathrm{d} x}  
- \frac{\gamma \bar{p}}{1-\gamma M_{x}^{2}} \left(\gamma M_{x}^{2} \beta - \alpha \right)\hat{u}_{x}
+ \gamma \bar{p} \frac{\mathrm{i} n}{r} \hat{u}_\theta{} = 0,
\end{equation}
\begin{equation}
\label{eq:momentum_x_LEEs}
\frac{1}{\bar{\rho}}\frac{\mathrm{d} \hat{p} }{\mathrm{d} x}
+ \frac{1}{\bar{\rho}} \frac{M_{x}^{2}}{1-\gamma M_{x}^{2}} \left(\beta - \alpha \right)  \hat{p}
+\bar{u}_x^{} \frac{\mathrm{d} \hat{u}_x^{}}{\mathrm{d} x} 
+  \left(\mathrm{i} \omega  +  \frac{\bar{u}_{x}}{1-\gamma M_{x}^{2}} \left(\beta - \alpha \right) \right) \hat{u}_{x} 
= \cancel{\bar{u}_x^{} \frac{\mathrm{d} \bar{u}_x^{}}{\mathrm{d} x} \hat{\sigma}},
\end{equation}
\begin{equation}
\label{eq:momentum_theta_LEEs}
\frac{1}{\bar{\rho}} \frac{\mathrm{i} n}{r}  \hat{p} + \bar{u}_x{} \frac{\mathrm{d} \hat{u}_\theta{} }{\mathrm{d} x}
+ \mathrm{i} \omega \hat{u}_\theta{} = 0,
\end{equation}
where $M_x=\bar{u}_x / \bar{c}$ indicates the axial Mach number and $\bar{c}$ denotes the local speed of sound. $\alpha$ and $\beta$ are normalised parameters defined as:
\begin{equation}
\label{eq:param}
\alpha  = \frac{1}{A}\frac{\mathrm{d} A}{\mathrm{d} x} = \frac{1}{h}\frac{\mathrm{d} h}{\mathrm{d} x}, \quad \beta  = \frac{1}{\overline{T}}\frac{\mathrm{d} \overline{T}}{\mathrm{d} x}.
\end{equation}

It should be noted that the imaginary part $k_i$ of the wave number $k$ is neglected in our previous work \citep{Li_JSV_2017,Li_JSV_2020},  as it is typically much smaller than the real part $k_r$, especially for  high frequency conditions. However, when there is sufficiently large heat source or acoustic energy sink within the thin annular combustion chamber, the absolute value of the imaginary part $k_i$, which typically corresponds to the growth rate of the wave, can be a large value that cannot be neglected. Thus, we have also accounted for  the imaginary part in the present work, and the wave number thus becomes a complex value and  can be expressed as:
\begin{equation}
\label{eq:kr_ki}
k = \frac{\omega}{\bar{c}} = k_{r} + \mathrm{i} k_{i} = \left( 1+\mathrm{i} \epsilon \right) k_{r}.
\end{equation}

Several assumptions are necessary for the next step, in which the LEEs are combined into a single wave equation for the acoustic pressure perturbation $\hat{p}$. Considering the fact that the inlet Mach number $M_{x,1}$ and value of $\epsilon$ $\left(\epsilon = k_{i} / k_{r} \right)$ are both small, the perturbation frequencies are relatively high and the nonuniformity remains small, all the assumptions adopted in the entire derivation are summarised as:
\begin{equation}
\label{eq:assumps}
|M_{x}^{3}| \ll 1, \quad
|\epsilon^{2}| \ll 1, \quad
\left|\frac{M_{x}}{k_{0}} \left( \beta - 2 \alpha \right) \epsilon \right| \ll 1,\quad
\left|\frac{1}{k_{0}^{2}} \beta \left(\beta - 2 \alpha \right) \right| \ll 1,\quad
\left|\frac{1}{k_{0}^{2}} \left(\frac{1}{2}\frac{\mathrm{d} \beta}{\mathrm{d} x} - \frac{\mathrm{d} \alpha}{\mathrm{d} x} \right) \right| \ll 1.
\end{equation}
%
%
%
%
%
With the assumptions of Eq.~\eqref{eq:assumps}, the LEEs can be combined into a single wave equation of the form:
\begin{equation}
\label{eq:wave_equation}
\begin{split}
\Big(1 - M_x^2  +  \frac{\mathrm{i} M_x}{ k_{r}} \left(\beta - 2 \alpha\right)\Big) \frac{\mathrm{d}^2 \hat{p}}{\mathrm{d}x^2} -
\Big( 2  \left(\mathrm{i} -\epsilon \right) & M_{x} k_{r} - 
\left(1 - M_x^2\right)\beta - \left(1 - \left(1+\gamma\right)  M_x^2\right)\alpha  \Big)\frac{\mathrm{d} \hat{p}}{\mathrm{d}x} +  \\
&
\Big(\left( 1 + 2\mathrm{i}\epsilon \right) k^{2}_{r} - \left( 1+\gamma \right) \mathrm{i} M_{x} k_{r} \beta - \frac{n^2}{r^2}(1 - \frac{\mathrm{i} M_x}{ k_r}  \beta) \Big) \hat{p} = 0.
\end{split}
\end{equation}

\subsection{WKB approximation and analytical solutions}
\label{subsec:2:2}

The modified WKB method used by Cummings \citep{Cummings_JSV_1977} provides an analytical way to solve complex acoustic fields. Supposing the pressure perturbation $\hat{p} (x)$ is determined by the amplitude factor $a(x)$ and phase factor $b(x)$, both  with real values,
\begin{equation}
\label{eq:WKB_p}
\hat{p}\left( x \right) = \mathscr{H} \exp\left(\int^x_{x_1} a + \mathrm{i} b\, \mathrm{d}x\right),\\
\end{equation}
where $\mathscr{H}$ is an arbitrary coefficient.
The wave equation Eq.~\eqref{eq:wave_equation} is divided into real and imaginary parts after substituting the expression of $\hat{p}$. For chambers without mean flow, $b$ is much larger than $a$. Under the assumptions of Eq.~\eqref{eq:assumps}, the approximate solutions of $b$ can be derived from the  real part of Eq.~\eqref{eq:WKB_p},
\begin{equation}
\label{eq:b}
b^\pm = \frac{M_x \mp \lambda}{1 - M_x^2} k_{r}, \quad
\lambda = \frac{k^*}{k_{r}} = \Big[ 1 - \big(1 - M_x^2 \big)\, \frac{n^2}{r^2 k_r^2} \Big]^{1/2},
\end{equation}
where $`+'$ and  $`-'$ stand for acoustic waves propagating downstream and upstream, respectively. The parameter $\lambda$ is calculated by the Mach number, the radius and wavenumbers, and $k^*$ is an equivalent wavenumber according to its expression. The frequency $\omega$ should be larger than the cut-off frequency $\omega_{c,n} = \max \left({|n|\bar{c} \sqrt{1-M_x^2}}/{r} \right)$ of the $n$th circumferential mode, where "max" means the maximum value of the entire domain. Only acoustic waves  with frequencies larger than the cut-off frequency can propagate in the annular ducts without attenuation.

Then the approximate solutions of $a^+$ or $a^-$ can be obtained by taking into account for the expressions of $b^+$ or $b^-$ and the solutions are:
\begin{equation}
\label{eq:a}
\begin{split}
a^\pm = -\frac{1}{2}\left( \alpha + \beta\right) & - \frac{1}{2 k^*}\frac{\mathrm{d} k^*}{\mathrm{d} x} 
\pm \left( \frac{1}{\lambda} \mp M_{x} +M^{2}_{x} \right) k_{i} \\
&\pm \left( \left(\frac{1-\gamma}{2 \lambda} - \lambda \right)\beta + \lambda \alpha \right)M_{x}
+ \left(\frac{1}{2}\beta - \frac{2-\gamma}{2}\alpha \right) M^{2}_{x}.
\end{split}
\end{equation}
%
%


 Considering the relation between the Mach number and its differential (Eq.~\eqref{eq:D_param}), 
\begin{equation}
\label{eq:dMx}
\frac{\mathrm{d} M_{x}}{\mathrm{d} x} \approx \left(\frac{1}{2} \beta - \alpha\right) M_{x},
\end{equation}
the equations can be further simplified:
%
%
\begin{equation}
\label{eq:a_int}
\begin{split}
\int a^{\pm} \mathrm{d} \zeta  = 
-\frac{1}{2}\ln \left(A \overline{T} k^{*}\right)   +  
\left(\frac{2-\gamma}{4}M^{2}_{x} \mp \lambda M_{x} \right) +  \int \left(\left(  \frac{\gamma}{4} \beta M^{2}_{x} \mp \frac{\gamma}{2 \lambda} \beta M_{x} \right) \pm
\left( \frac{1}{\lambda} \mp M_{x} + M^{2}_{x} \right) k_{i} \right) \mathrm{d} \zeta. 
\end{split}
\end{equation}

According to the expressions for $a^{\pm}$ and $b^{\pm}$, the general analytical solutions of the acoustic pressure perturbation $\hat{p}$ as  functions of  $x$  and  $\omega$ can be obtained:
\begin{equation}
\label{eq:wave_hat_p}
\hat{p} \left(x,\omega\right) = {\mathcal{C}}^+ {\mathcal{P}}^+ \left(x,\omega\right)
+ {\mathcal{C}}^- {\mathcal{P}}^- \left(x,\omega\right),
\end{equation}
where,
\begin{equation}
\label{eq:P+-}
\begin{split}
{\mathcal{P}}^ \pm \left(x,\omega\right) = &
\left( \frac{A_{1} \overline{T}_{1} k^{*}_{1}}{A \overline{T} k^{*}} \right)^{1/2}
 \cdot  \frac{\mathrm{exp}\left( \pm \lambda_{1} M_{x,1} - \frac{2-\gamma}{4} M_{x,1}^2  \right)}
{\mathrm{exp}\left( \pm \lambda M_{x} - \frac{2-\gamma}{4} M_{x}^2 \right)}  \\
& \; \cdot \mathrm{exp}\left[ \int_{x_1}^x \left(\left( \mp \frac{\gamma}{2 \lambda} \beta M_{x} + \frac{\gamma}{4} \beta M^{2}_{x} \right) 
 \pm \left( \frac{1}{\lambda} \mp M_{x} + M^{2}_{x} \right) k_{i} 
+ \frac{M_x \mp \lambda}{1 - M_x^2} \cdot \mathrm{i} k_{r} \right) \mathrm{d} \zeta \right] .
\end{split}
\end{equation}

$\mathcal{C}^+$ and $\mathcal{C}^-$ are two coefficients which can be determined from given initial and boundary conditions. The expression for $\mathcal{P}^{\pm}$ indicates that the acoustic waves in the duct have
strengths proportional to $\left(A \overline{T} k^{*}\right)^{1/2}$, which agrees with the results in Ref. \citep{Subrahmanyam_JSV_2001}. The comparisons of the analytical solutions to other references are illustrated in detail in  \ref{sec:Appendix_com_ana}.

The analytical solution of circumferential velocity perturbation can be obtained by substituting $\hat{p}$ into the circumferential momentum conservation equation (Eq.~\eqref{eq:momentum_theta_LEEs}) with the boundary condition, e.g., zero velocity at the inlet,
\begin{equation}
\label{eq:hat_u_theta}
\hat{u}_{\theta} = 
\exp \left( -\int_{x_1}^x \frac{\mathrm{i} \omega}{\bar{u}_x} \mathrm{d}x \right) 
\int_{x_1}^x \left[ -\frac{\mathrm{i} n}{r}\frac{\hat{p} } {\bar{\rho} \bar{u}_x} \mathrm{exp} \left(  \int_{\zeta_1}^\zeta \frac{\mathrm{i} \omega}{\bar{u}_x}\mathrm{d} \eta \right) \right] \mathrm{d}\zeta .
\end{equation}

Combining Eqs.~\eqref{eq:energy_LEEs} and \eqref{eq:momentum_x_LEEs}, the analytical solution of the axial velocity perturbation $\hat{u}_{x}$ can be obtained,
\begin{equation}
\label{eq:hat_u_x}
\hat{u}_{x}  = \frac{1}{\bar{\rho} \bar{c} \mathcal{D}} \Bigg[ M_{x} \Big[\frac{M_{x}}{1 - \gamma M^{2}_{x}} \left( \left( \gamma -1 \right) \beta + \left( 1 - \gamma^{2} M^{2}_{x} \right) \alpha \right)
+ \left( \mathrm{i} - \epsilon \right) k_{r}
\Big] \hat{p} 
-\left( 1-M^{2}_{x} \right) \frac{\mathrm{d} \hat{p} }{\mathrm{d} x} \Bigg] + \frac{\mathrm{i} n}{r} \frac{M_{x}}{\mathcal{D}} \hat{u}_{\theta}.
\end{equation}
According to the expressions for $\hat{p}$ and $\hat{u}_\theta$, the solution of $\hat{u}_{x}$ is,
\begin{equation}
\label{eq:hat_ux_B_D}
\hat{u}_{x}  =
\frac{\mathcal{B}^+ }{\mathcal {D}}   \cdot \frac{{\mathcal{C}}^+ {\mathcal{P}}^+}{\bar{\rho}\bar{c} }
- \frac{\mathcal{B}^- }{\mathcal {D}}   \cdot \frac{{\mathcal{C}}^- {\mathcal{P}}^-}{\bar{\rho}\bar{c}}
+ \dfrac{\mathrm{i} n}{r} \dfrac{M_x }{\mathcal {D}}\cdot \hat{u}_\theta,
\end{equation}
\begin{equation}
\label{eq:ux_B}
\mathcal{B}^\pm =  \left( \mathrm{i} \lambda - \frac{1}{\lambda} \epsilon \right) k_{r} \pm \left(\frac{1}{2} \mp \lambda M_{x} + \frac{3-\gamma}{2} M^{2}_{x} \right) \alpha  \pm   \left[\left( \frac{1}{2} - \frac{1}{4 \lambda^{2}}\right) \pm \left( \lambda +\frac{\gamma - 1}{2 \lambda} \right) M_{x} + \frac{4 \gamma - 7}{4} M^{2}_{x} \right] \beta,
\end{equation}
\begin{equation}
\label{eq:ux_D}
\mathcal{D} = \left( \mathrm{i} - \epsilon \right) k_{r} + \frac{M_{x}}{1 - \gamma M^{2}_{x}} \left[ \left(1 + \gamma M^{2}_{x}\right) \beta - 2 \alpha \right] 
\end{equation}

Eqs.~\eqref{eq:wave_hat_p}-\eqref{eq:ux_D} are the analytical solutions for the acoustic field in a thin annular chamber with axial mean flow and an axially-varying annular gap height and mean temperature. These solutions are valid for cases satisfying the conditions in Eq.~\eqref{eq:assumps} and will be further validated in the following sections. To fully simplify the analytical solutions, stronger assumptions than Eq.~\eqref{eq:assumps} are also considered. For low Mach number flow (e.g., $M_{x}<0.1$), the analytical solutions can be further simplified as follows and these solutions are calculated together with Eqs.~\eqref{eq:wave_hat_p}-\eqref{eq:ux_D} in the results.
%
%
\begin{equation}
\label{eq:P+-_L}
\begin{split}
{\mathcal{P}}^ \pm_{L}  = 
\left( \frac{A_{1} \overline{T}_{1} k^{*}_{1}}{A \overline{T} k^{*}} \right)^{1/2}
   \frac{\mathrm{exp}\left( \pm \lambda_{1} M_{x,1} \right)}
{\mathrm{exp}\left( \pm \lambda M_{x} \right)}  
\cdot \mathrm{exp}\left[ \int_{x_1}^x \left( \mp \frac{\gamma}{2 \lambda} \beta M_{x} 
 \pm \left( \frac{1}{\lambda} \mp M_{x} \right) k_{i} 
+ \frac{M_x \mp \lambda}{1 - M_x^2} \cdot \mathrm{i} k_{r} \right) \mathrm{d} \zeta \right],
\end{split}
\end{equation}
\begin{equation}
\label{eq:ux_B_L}
\mathcal{B}^\pm_{L} =  \left( \mathrm{i} \lambda - \frac{1}{\lambda} \epsilon \right) k_{r} \pm \left(\frac{1}{2} \mp \lambda M_{x} \right) \alpha  \pm   \left[\left( \frac{1}{2} - \frac{1}{4 \lambda^{2}}\right) \pm \frac{\gamma + 1}{2 \lambda} M_{x} \right] \beta,
\end{equation}
\begin{equation}
\label{eq:ux_D_L}
\mathcal{D}_{L} = \left( \mathrm{i} - \epsilon \right) k_{r} + M_{x} \beta -2 M_{x} \alpha.
\end{equation}

\section{Validation configurations}
\label{sec:3}

In the derivation of the wave equation (Eq.~\eqref{eq:wave_equation}) and the final analytical solutions (Eqs.~\eqref{eq:wave_hat_p}-\eqref{eq:ux_D}), some terms are neglected under appropriate assumptions in Eqs.~\eqref{eq:momentum_x_LEEs}, \eqref{eq:assumps} and \eqref{eq:dMx}. The validation of the analytical solution  is performed by considering the perturbation terms corresponding to the acoustic field in a thin annular chamber, e.g., the pressure perturbation $\hat{p}$, axial velocity perturbation $\hat{u}_x$ and circumferential  velocity perturbation $\hat{u}_\theta$. Following our previous works \citep{Li_JSV_2017,Li_JSV_2020}, we also use the normalised perturbation terms (also called transfer functions) for the comparisons, expressed as: 
\begin{equation}
\begin{split}
\label{eq:TF}
\mathcal{F}_{p}(x, \omega) = \frac{\hat{p}(x, \omega)}{\hat{p}(x_1, \omega)}, \quad
\mathcal{F}_{u_{x}}(x, \omega) = \frac{\hat{u}_{x}(x, \omega)}{\hat{u}_{x}(x_1, \omega)}   \quad \textrm{and} \quad \mathcal{F}_{u_{\theta}}(x, \omega) = \frac{\hat{u}_{\theta}(x, \omega)}{\hat{u}_{x}(x_1, \omega)}.
\end{split}
\end{equation}

\begin{figure}[!h]
	\centering
	\subfigure
	{
		\includegraphics[width=6cm]{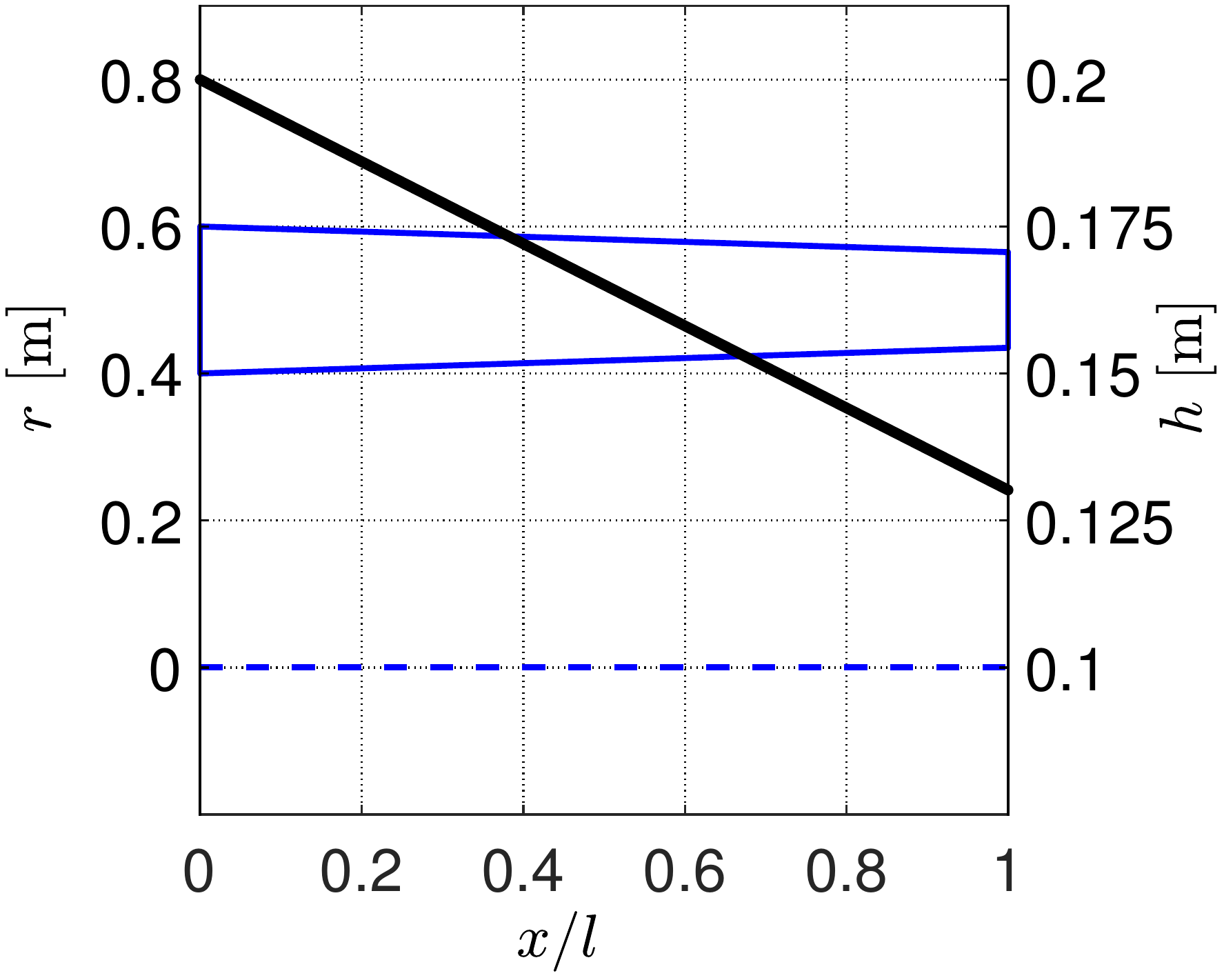}
		\label{Fig:Plot_h}
	}
	\vspace*{-0pt}
	\hspace*{10pt}
	\caption{The geometry profiles of the thin annular combustion chamber along the axial location corresponding to Table~\ref{table_parameters}. Thin lines represent  the geometry profiles and  the dashed one stands for the symmetry axis. The thick line stands for the height of the annular gap, whose value is marked in the right axis. }
	
	\label{Fig:Plot_h_Ta} 
	\vspace*{00pt}
\end{figure}

The acoustic waves along the circumferential direction are uniform due to the circumferential homogeneous flow distribution assumption. It is thus possible to use the axial distributions of acoustic waves at the zero circumferential angle for the analysis of the acoustic field. Furthermore, the angular frequency $\omega$ is a complex number in this work and is expressed as  $\omega = 2 \pi \bar{c}_1/l H_e + \mathrm{i} G_r$, where $H_e$ is a normalised frequency (also called the Helmholtz number), and $G_r$ is the modal growth rate. 
The numerical results against which the analytical solution is compared are calculated by simulating the three LEEs (Eqs.~\eqref{eq:energy_LEEs}-\eqref{eq:momentum_theta_LEEs}) from which  the proposed  analytical solutions were originally derived. 
A fourth order finite difference scheme is used for the numerical method, with the duct divided into a $6000$ points uniform grid. 
The parameters used for the analysis in the following sections are listed in Table~\ref{table_parameters} unless otherwise stated. 
The geometry profiles and the gap height $h$ along the annular duct are shown in Fig.~\ref{Fig:Plot_h_Ta}.
The inlet of the duct is an anechoic boundary and the outlet is open to the atmosphere $(\hat{p}_2=0)$. 
A pressure perturbation with amplitude $1000$ Pa is prescribed at the inlet. 

\begin{table*}[ht]
	\caption{Parameters used in the validation. }
	\centering
	\label{table_parameters}
		\begin{tabular}{cccccccc}
			\hline\hline
			$l$ [m]   &$r_1$ [m]   &$\bar{p}_1$ [Pa]   &$M_{x,1}$ [-]   &$He$ [-]   &$Gr$ [$s^{-1}$]   &$\omega$ [$s^{-1}$]   &$\gamma$ [-]\\
			\hline
			1   &0.5   &101325   &0.1   &0.5   &100   &2519+100$\mathrm{i}$    &1.4\\
			\hline
			$h_{1}$ [m]   &$h_{2}$ [m]   &$\alpha_h$ [$^\circ$]   &$\overline{T}_{1}$ [K]   &$\overline{T}_{2}$ [K]   &$\alpha_{\overline{T}}$ [K/m]   &$R_g$ [J K $^{-1}$ kg$^{-1}$]\\
			\hline
			0.2   &0.13   &-2   &1600   &1000   &-600   &287 \\
			\hline\hline
		\end{tabular}
\end{table*}
%

\section{Results and discussions}
\label{sec:4}
\subsection{Comparisons of transfer functions for  different flow and modal conditions}
\label{subsec:3:1}

Comparisons of the  normalised perturbation terms (transfer functions) along the thin annular chamber calculated by the analytical solutions and numerical results are then conducted for different flow conditions, circumferential  wave number and modal growth rates. Results of Eqs.~\eqref{eq:P+-_L}-\eqref{eq:ux_D_L} are also calculated to figure out the applicability of simplified analytical solutions.

\subsubsection{Comparisons for different flow Mach numbers $M_{x}$}
\label{subsubsec:3:1:1}

\begin{figure}[!h]
	\hspace*{13pt}
	\subfigure
	{
		\includegraphics[height=7cm]{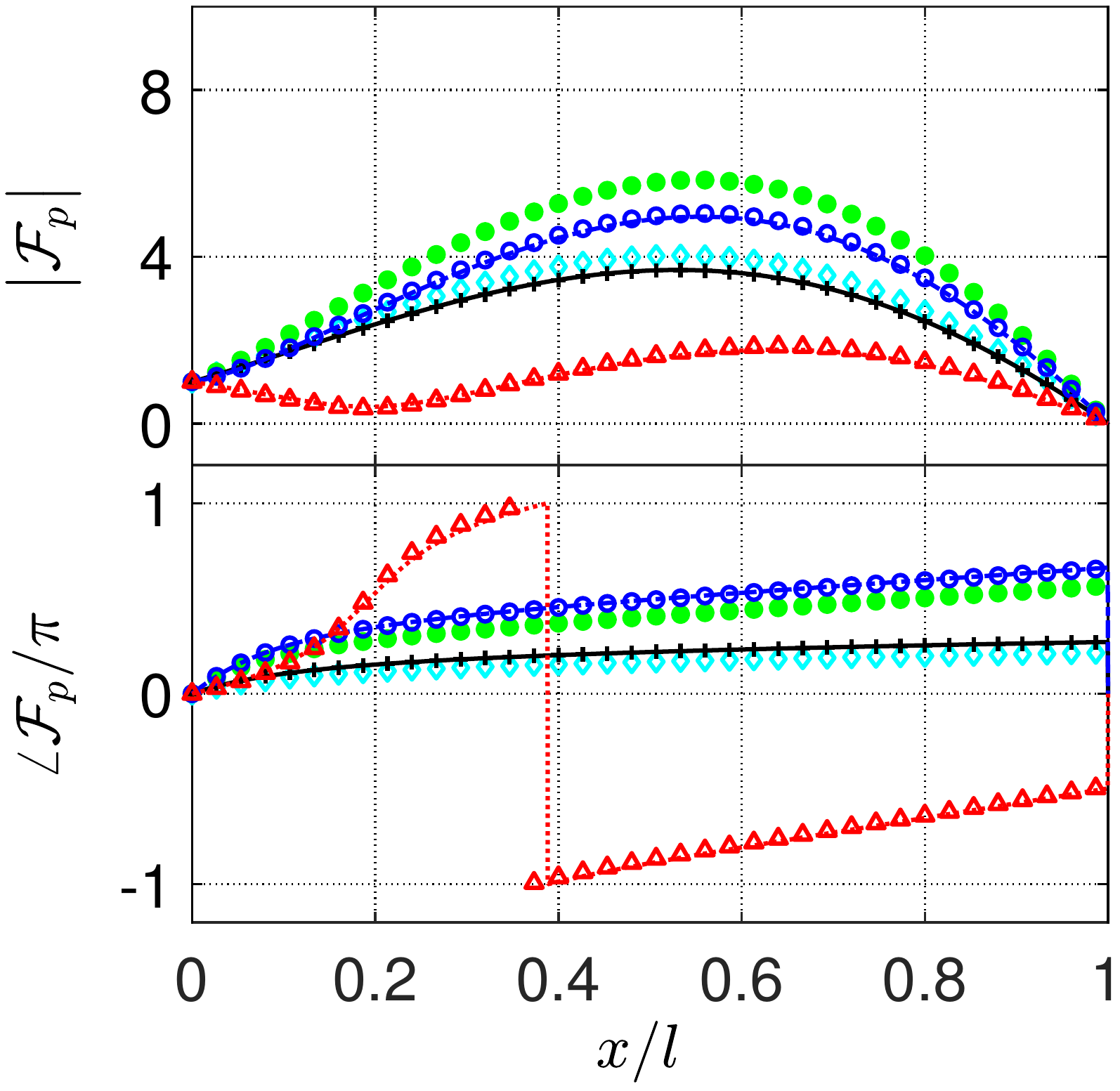}
		\label{Fig:M3_p}
	}
	\put (-200,180) {\normalsize$\displaystyle(a)$}
	\vspace*{0pt}
	\hspace*{10pt}
	\subfigure
	{
		\includegraphics[height=7cm]{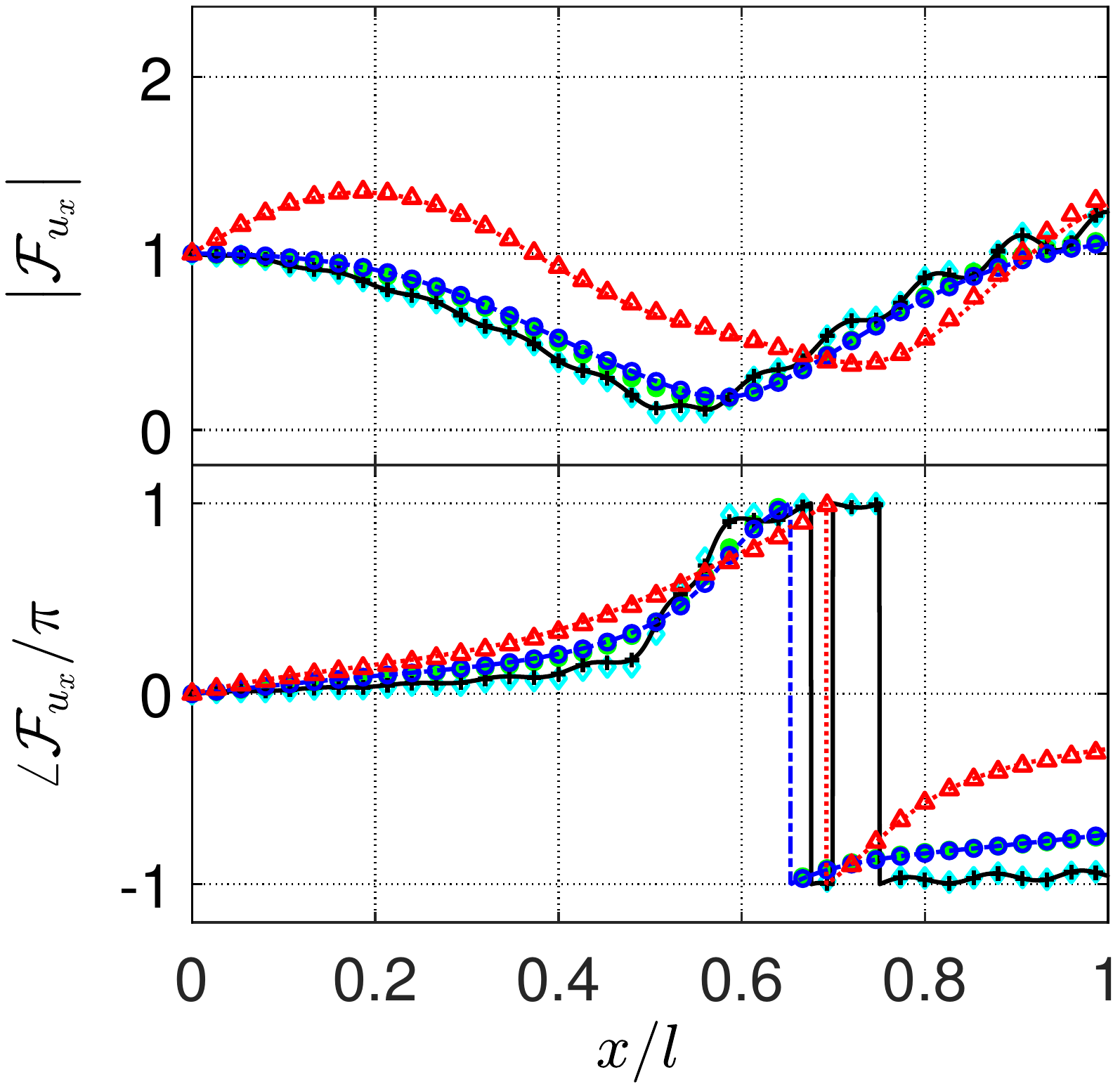}
		\label{Fig:M3_ux}
	}
	\put (-200,180) {\normalsize$\displaystyle(b)$}
	\vspace*{-0pt}\\
	\hspace*{13pt}
	\subfigure
	{
		\includegraphics[height=7cm]{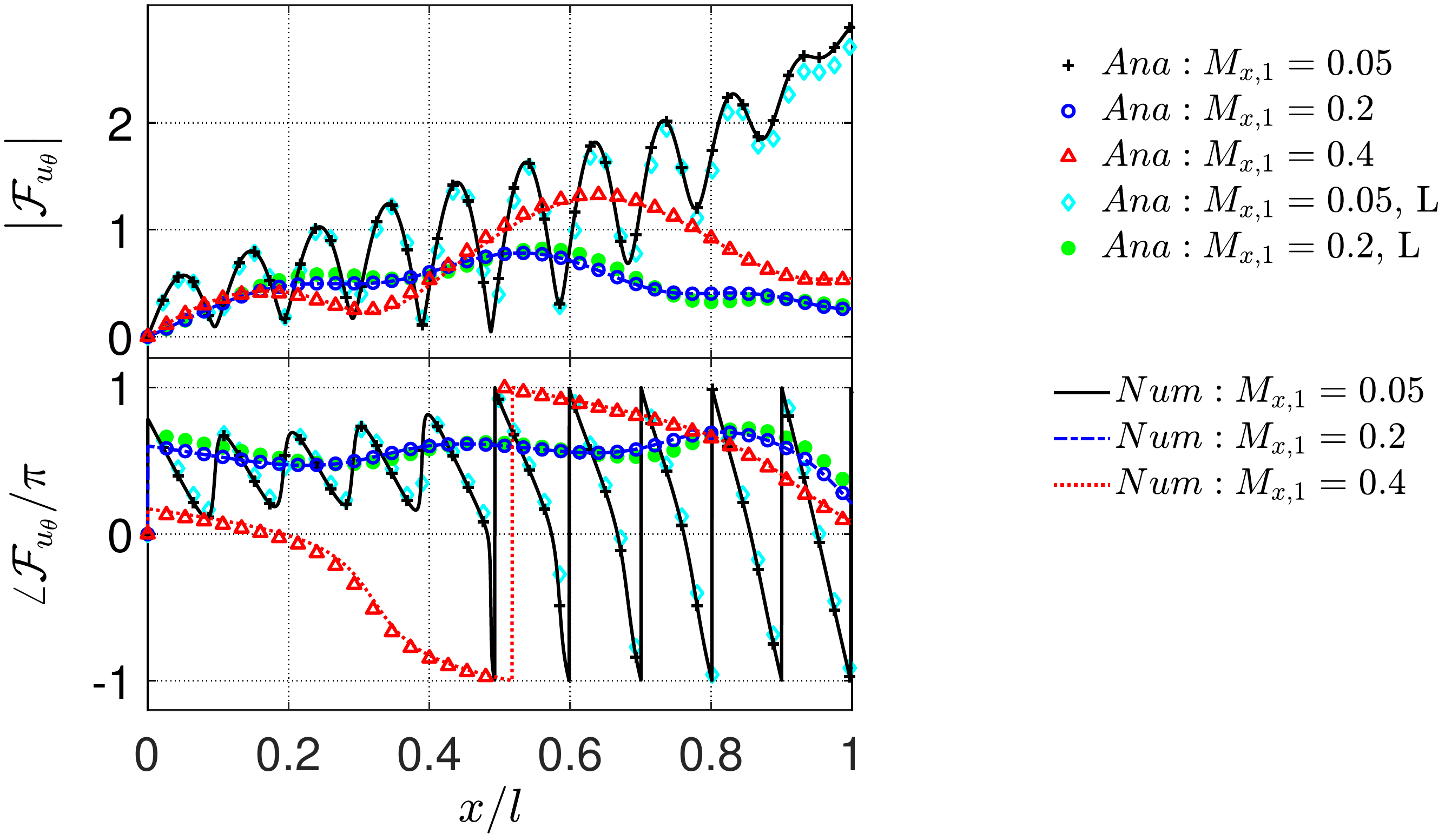}
		\put (-333,180) {\normalsize$\displaystyle(c)$}
		\label{Fig:M3_uth}
	}
	\vspace*{0pt}
	\caption{Axial distributions of  $\mathcal{F}_p$, $\mathcal{F}_{u_x}$ and $\mathcal{F}_{u_\theta}$ calculated by the analytical and numerical methods for different inlet flow Mach numbers ($M_{x,1} = 0.05, 0.2$ and $0.4$). The numerical ($Num$) and analytical ($Ana$) solutions are represented by lines and symbols, respectively. Herein,  $n = 1$. Simplified analytical results are marked by diamond and solid circle.}
	
	\label{Fig:M_005_02_04} 
	\vspace*{00pt}
\end{figure}

Figure~\ref{Fig:M_005_02_04} shows the transfer functions of acoustic perturbations ($\mathcal{F}_p$, $\mathcal{F}_{u_x}$ and $\mathcal{F}_{u_\theta}$) as functions of  the axial distance $x/l$ for three inlet flow Mach numbers ($M_{x,1} = 0.05, 0.2$ and  $0.4$). The solutions from the two methods collapse for the three flow conditions, even when the flow Mach number $M_{x}$ equals $0.4$ at the inlet (the Mach number at the outlet equals 0.4785), indicating that this solution is valid for a large range of flow velocity conditions.  The differences among these transfer functions for the pressure and velocity perturbations are obvious for different inlet Mach numbers.  With increasing the inlet flow Mach number, the phase change of $\mathcal{F}_p$ and $\mathcal{F}_{u_x}$ also increase, while that of $\mathcal{F}_{u_\theta}$ decreases. 
Results of simplified analytical solutions when $M_{x,1} = 0.05$ also have good accuracy, which means that the simplified equations are suitable for cases with low inlet Mach numbers.

\subsubsection{Comparisons for different circumferential wavenumbers $n$}
\label{subsubsec:3:1:2}

\begin{figure}[!h]
	\hspace*{13pt}
	\subfigure
	{
		\includegraphics[height=7cm]{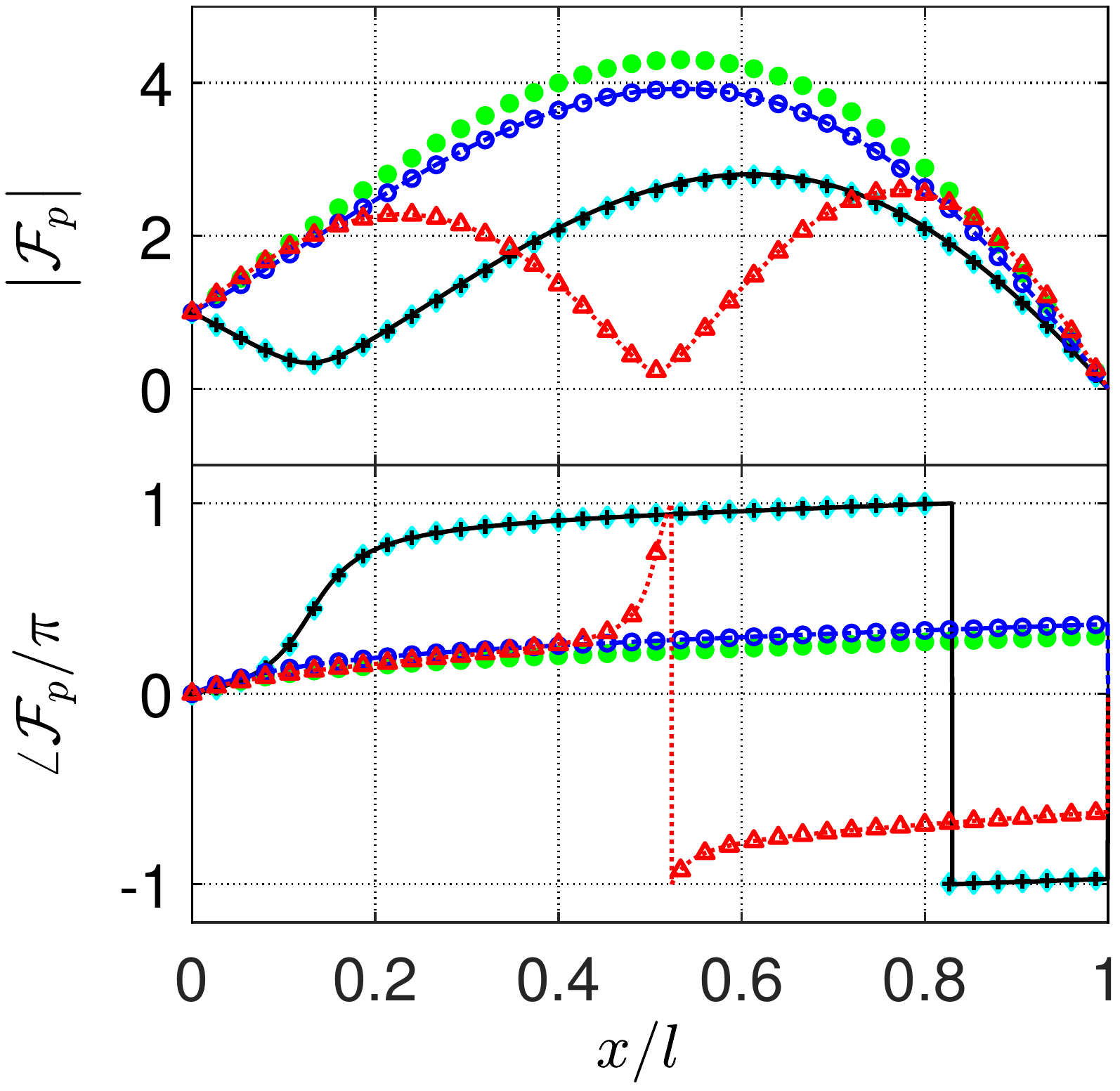}
		\label{Fig:n3_p}
	}
	\put (-200,180) {\normalsize$\displaystyle(a)$}
	\vspace*{-0pt}
	\hspace*{10pt}
	\subfigure
	{
		\includegraphics[height=7cm]{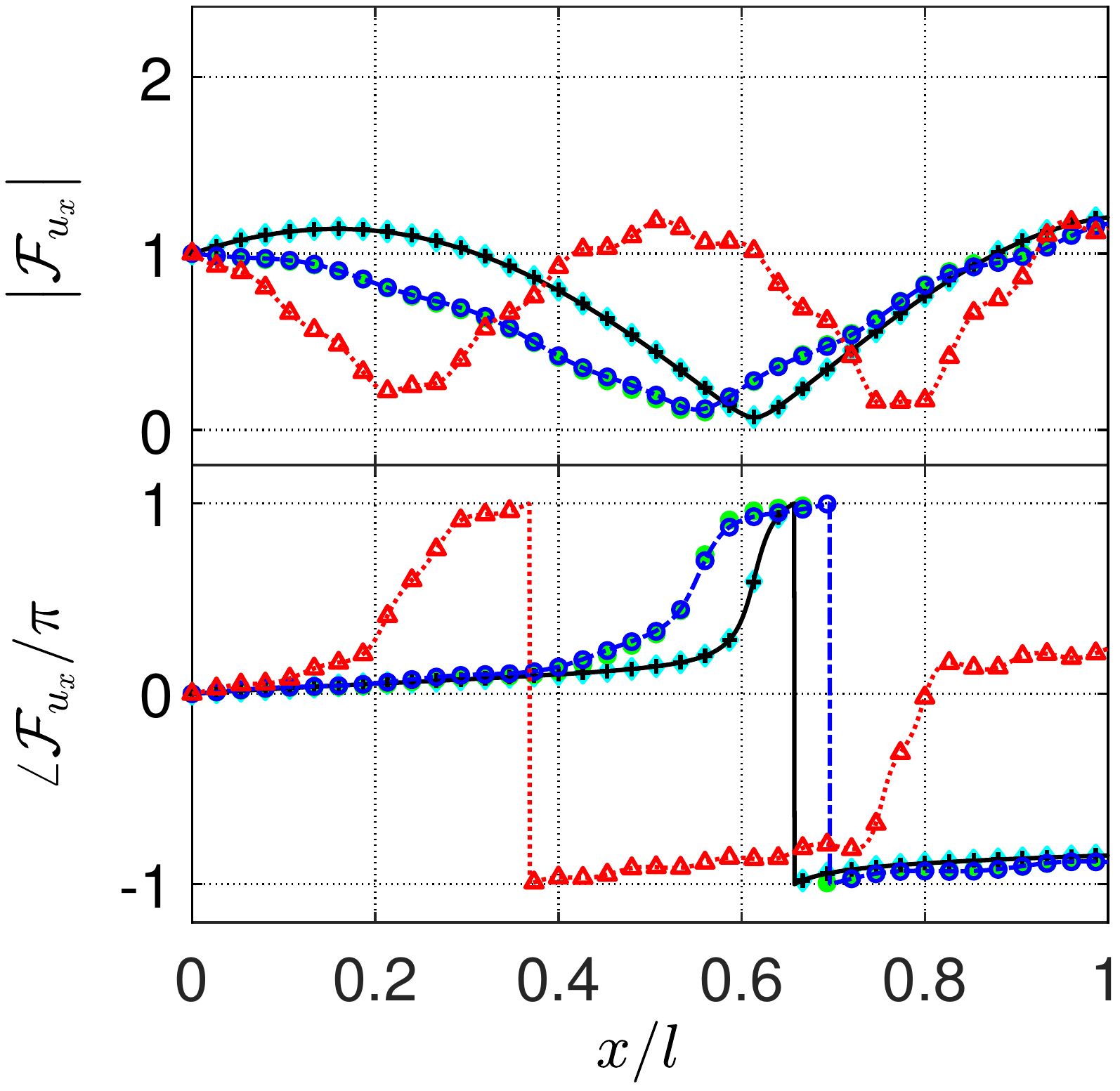}
		\label{Fig:n3_ux}
	}
	\put (-200,180) {\normalsize$\displaystyle(b)$}
	\vspace*{-0pt}\\
	\hspace*{13pt}
	\subfigure
	{
		\includegraphics[height=7cm]{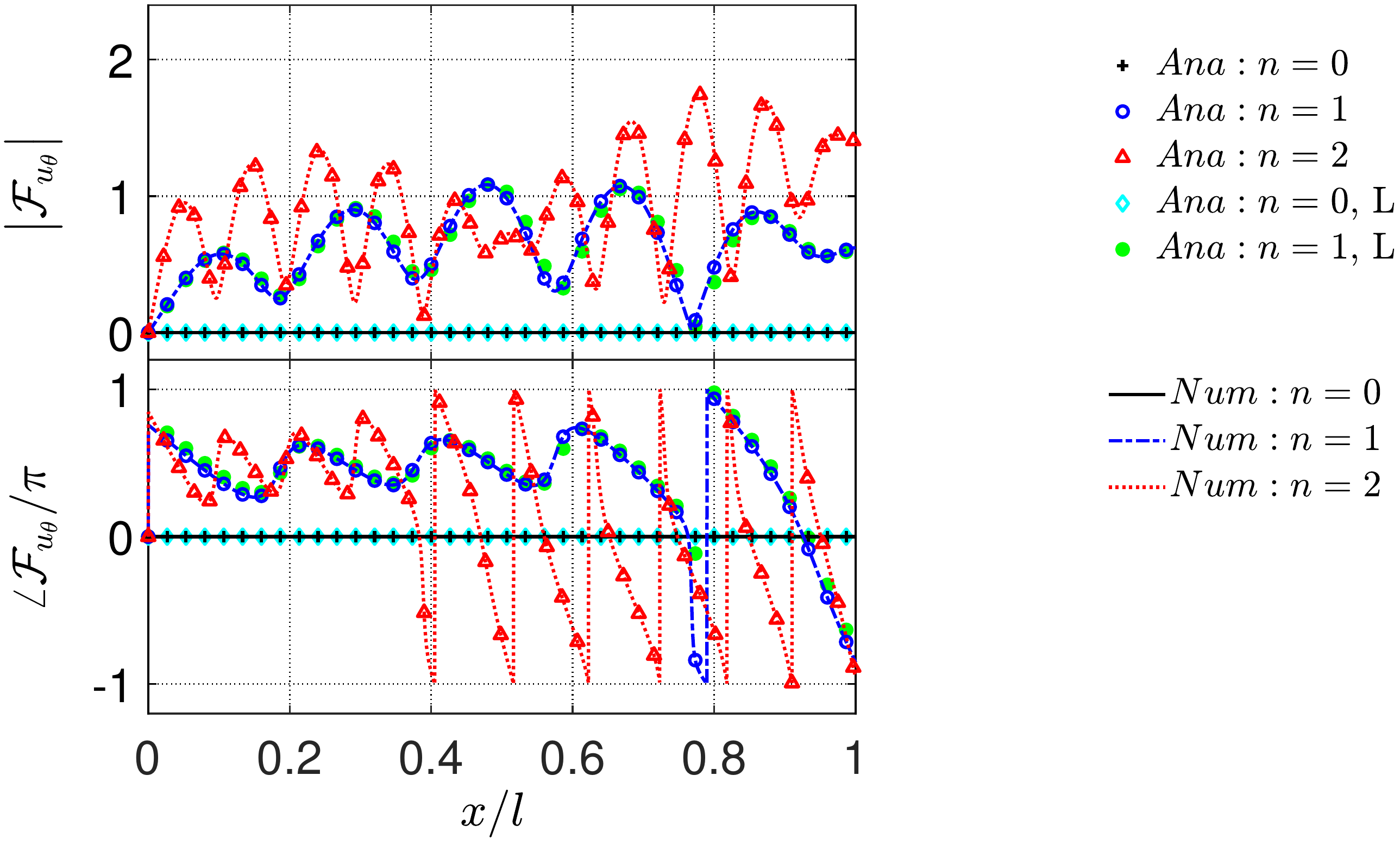}
		\put (-323,180) {\normalsize$\displaystyle(c)$}
		\label{Fig:n3_uth}
	}
	\vspace*{-0pt}
	\caption{Comparisons of the axial distributions of  $\mathcal{F}_p$, $\mathcal{F}_{u_x}$ and $\mathcal{F}_{u_\theta}$ calculated by the analytical and numerical methods for  different circumferential wavenumbers ($n = 0, 1$ and $2$). The numerical ($Num$) and analytical ($Ana$) solutions are represented by lines and symbols, respectively. Herein, $M_{x} = 0.1$. Simplified analytical results are marked by diamond and solid circle.}
	
	\label{Fig:n_0_1_2} 
	\vspace*{00pt}
\end{figure}

Effects of the circumferential wavenumber on the performance of the analytical solutions are now considered. It should be noted that the present analytical solution is valid for a relatively large frequency condition. 
For a given  wavenumber $k$ (or perturbation frequency $\omega$), the axial component decreases with increasing the circumferential wavenumber $n$.  The error in the prediction from the analytical solution  for a larger value of $n$ thus may become larger than that for a smaller value of $n$ when the wavenumber $k$ is fixed.  
Fig.~\ref{Fig:n_0_1_2} shows  the comparisons of transfer functions of acoustic perturbations ($\mathcal{F}_p$, $\mathcal{F}_{u_x}$ and $\mathcal{F}_{u_\theta}$) as functions of  the axial distance $x/l$ for three  circumferential wavenumbers $n = 0, 1$ and 2. The two results still collapse, validating the proposed analytical solutions.  
The phases of  $\mathcal{F}_{u_\theta}$ for the three cases have similar changing speed along the duct due to the same inlet Mach number and boundary conditions; the differences in the amplitudes and phases of $\mathcal{F}_p$ and $\mathcal{F}_{u_x}$ remain large.
When $n$ = 0, there is no circumferential mode and $\mathcal{F}_{u_\theta}$ thus equals zero.
Curves of $\mathcal{F}_{u_x}$ with large circumferential wavenumbers have many wraps (case of $n = 2$ as shown in Fig.~\ref{Fig:n3_ux}) because of the rapid changing of circumferential velocity perturbation along the axis. Results from simplified analytical solutions are similar to the two former methods. But large errors occur in the final results with large circumferential wavenumbers.

\subsubsection{Comparisons for different  modal growth rates}
\label{subsubsec:3:1:3}

%
\begin{figure}[!h]
	\hspace*{13pt}
	\subfigure
	{
		\includegraphics[height=7cm]{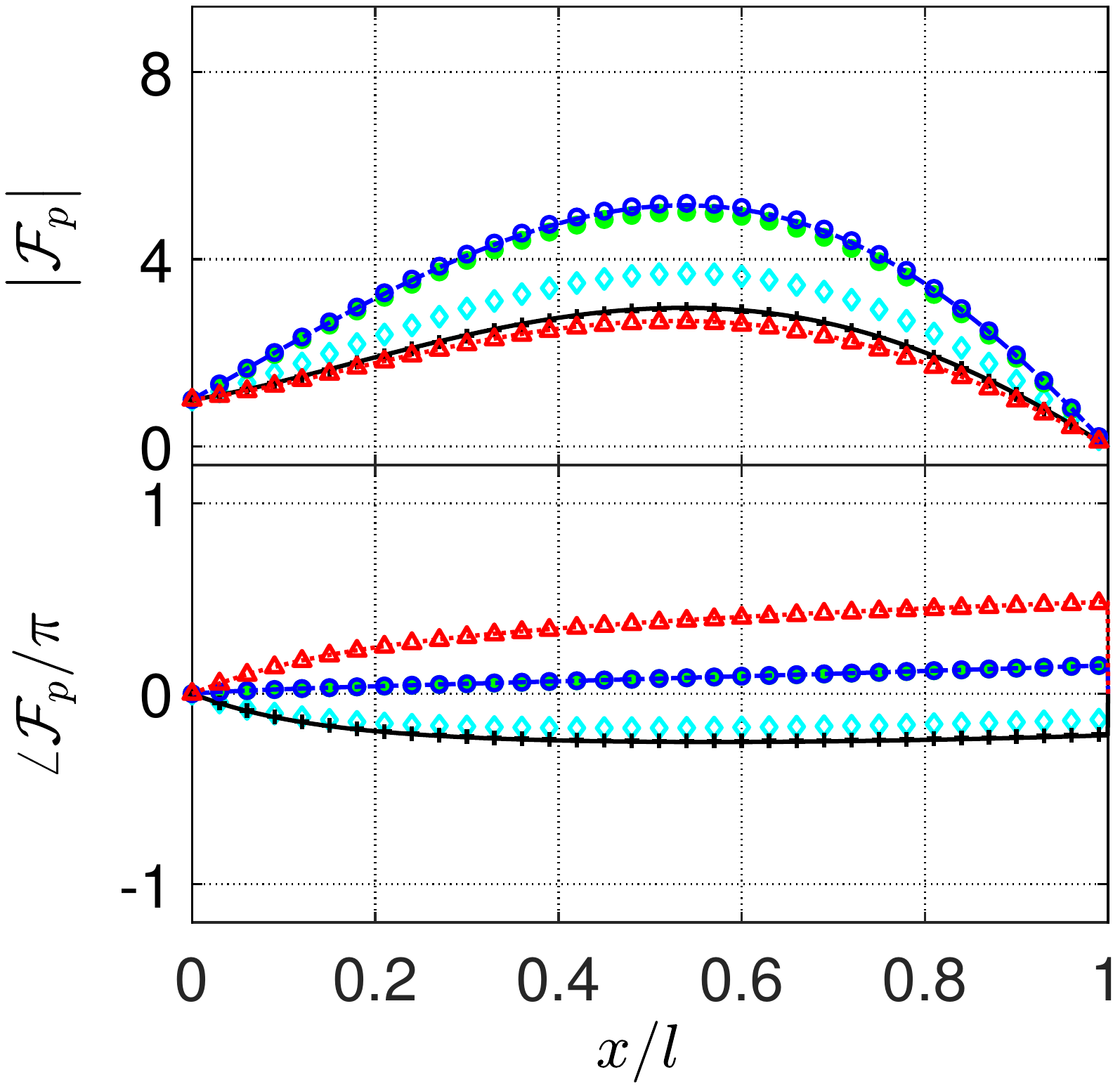}
		\label{Fig:G3_p}
	}
	\put (-200,180) {\normalsize$\displaystyle(a)$}
	\vspace*{-0pt}
	\hspace*{10pt}
	\subfigure
	{
		\includegraphics[height=7cm]{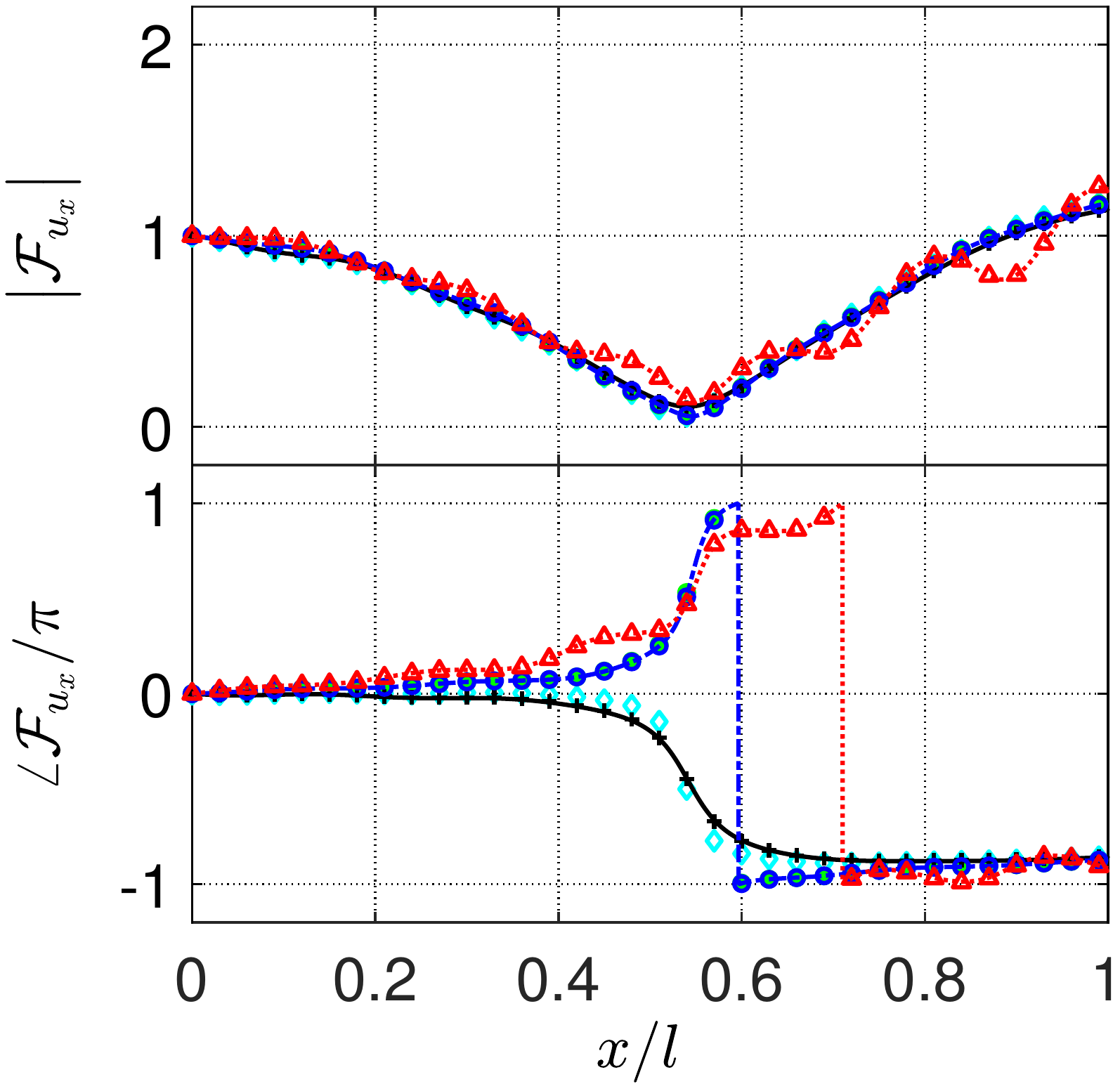}
		\label{Fig:G3_ux}
	}
	\put (-200,180) {\normalsize$\displaystyle(b)$}
	\vspace*{-0pt}\\
	\hspace*{13pt}
	\subfigure
	{
		\includegraphics[height=7cm]{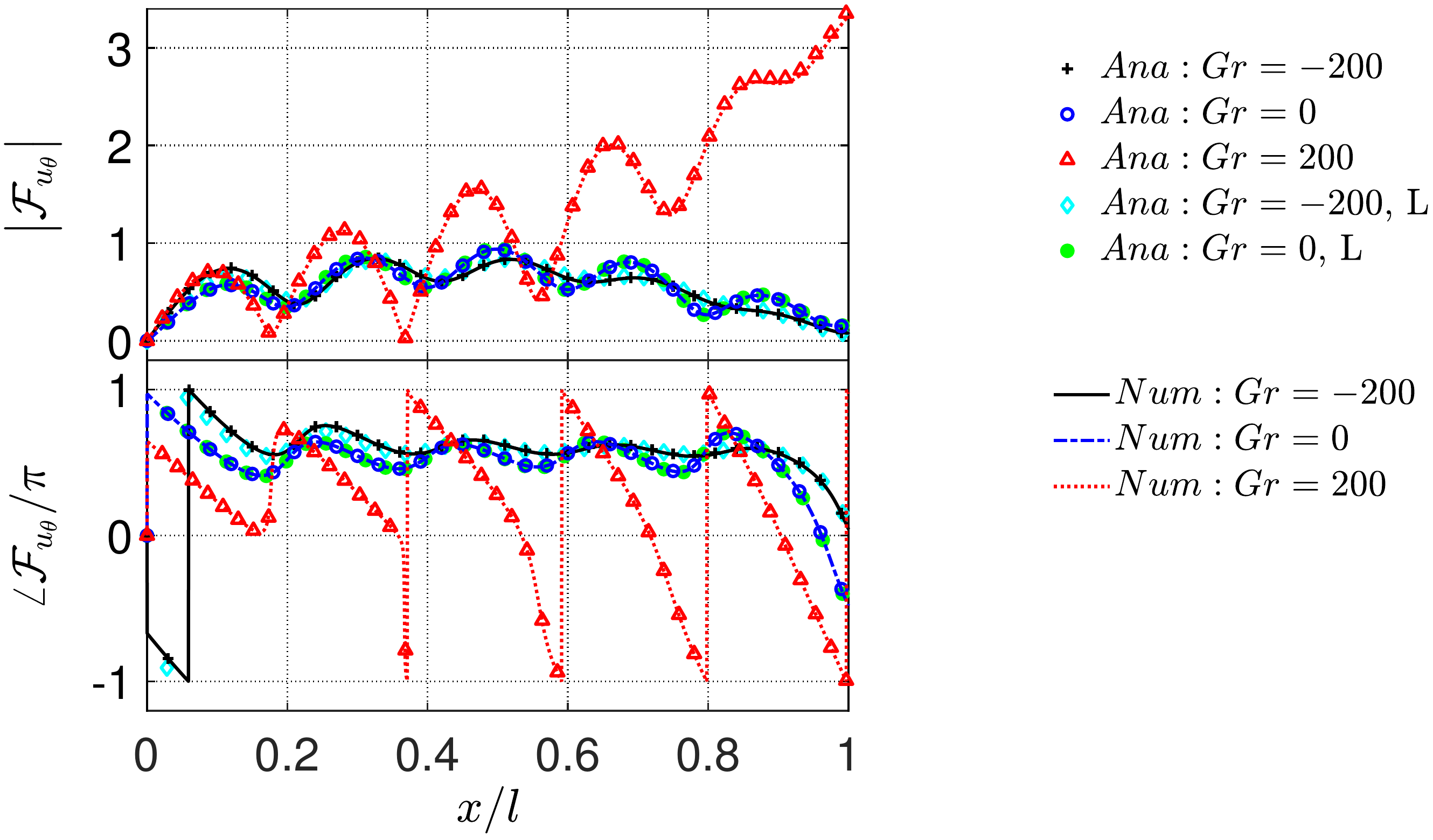}
		\put (-333,180) {\normalsize$\displaystyle(c)$}
		\label{Fig:G3_uth}
	}
	\vspace*{-0pt}
	\caption{Comparisons of the axial distributions of  $\mathcal{F}_p$, $\mathcal{F}_{u_x}$ and $\mathcal{F}_{u_\theta}$ calculated by the analytical and numerical methods for  different modal growth rates ($Gr = -200, 0$ and 200 $s^{-1}$). The numerical ($Num$) and analytical ($Ana$) solutions are represented by lines and symbols, respectively. Herein, $M_{x} = 0.1$ and  $n = 1$. Simplified analytical results are marked by diamond and solid circle.}
	
	\label{Fig:Gr_-2_0_2}
	\vspace*{00pt}
\end{figure}
%

In this section, a pressure perturbation with a complex frequency $\omega = 2\pi \bar{c}_1/l H_e + \mathrm{i} G_r$ is prescribed at the inlet of the thin annular duct to examine the performance of the present analytical solution when the absolute value of the growth rate  is large. 
Fig.~\ref{Fig:Gr_-2_0_2} shows the comparisons for three modal growth rates $Gr = -200, 0$ and 200 s$^{-1}$.
The predictions from the analytical solutions still match the numerical results. $\mathcal{F}_p$ and  $\mathcal{F}_{u_x}$ have similar amplitude trajectories  as shown in   Fig.~\ref{Fig:G3_p}-\ref{Fig:G3_ux}, but the differences among  the trajectories of transfer function phases  are large. 
The differences among the transfer functions of $\hat{u}_\theta$ are both large for the amplitudes and phases, as shown in Fig.~\ref{Fig:G3_uth}. 
Two analytical results match well when the imaginary part of $k$ equals zero but differences occur when $Gr = -200$ s$^{-1}$.


\subsection{Contour maps of error coefficients}
\label{subsec:3:2}

\begin{figure}[!h]
	\centering
	\subfigure
	{
		\includegraphics[width=7cm]{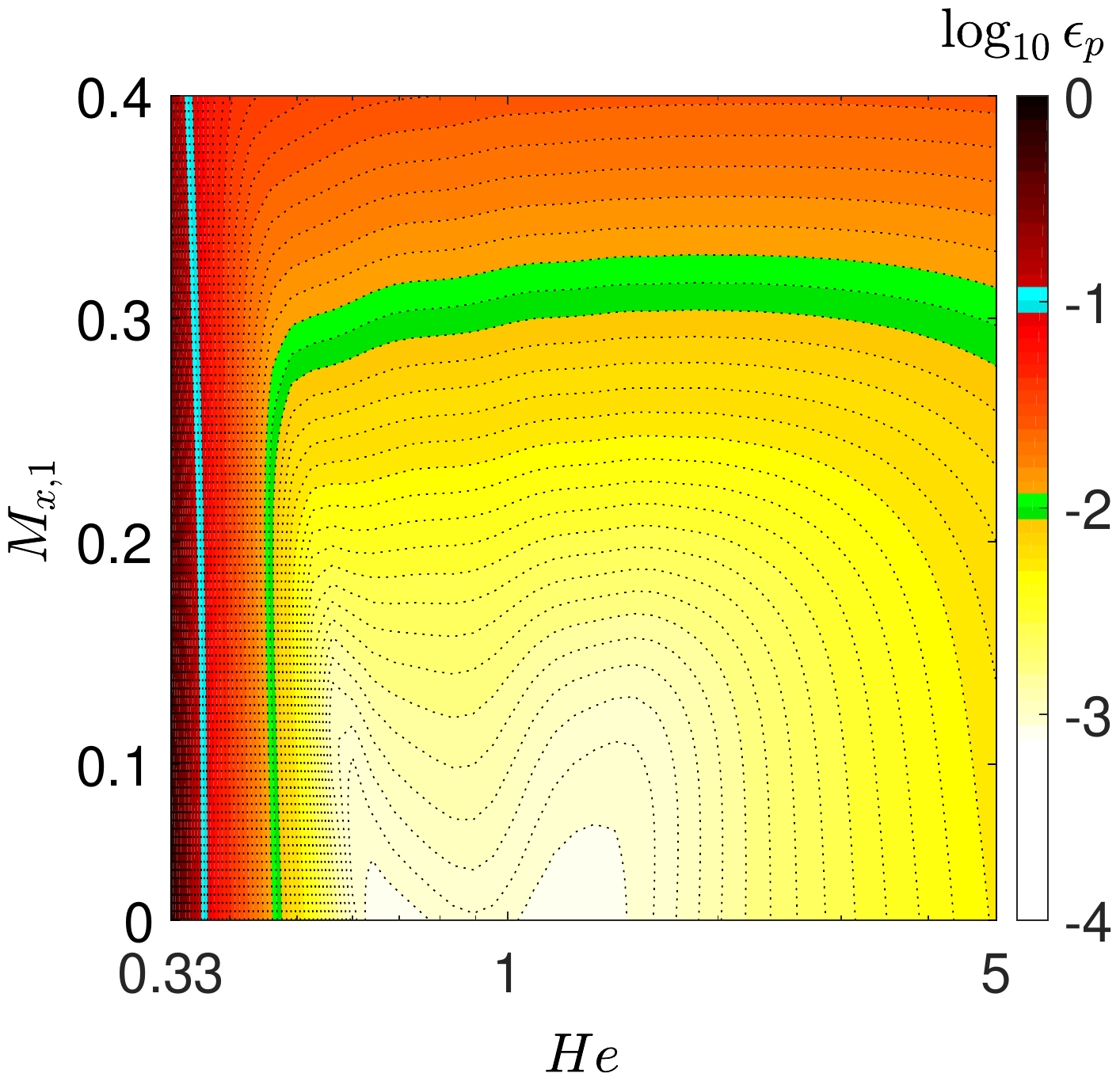}
		\label{Fig:Contour_Mx1_He_p}
	}
	\put (-205,170) {\normalsize$\displaystyle(a)$}
	\vspace*{-0pt}
	\hspace*{30pt}
	\subfigure
	{
		\includegraphics[width=7cm]{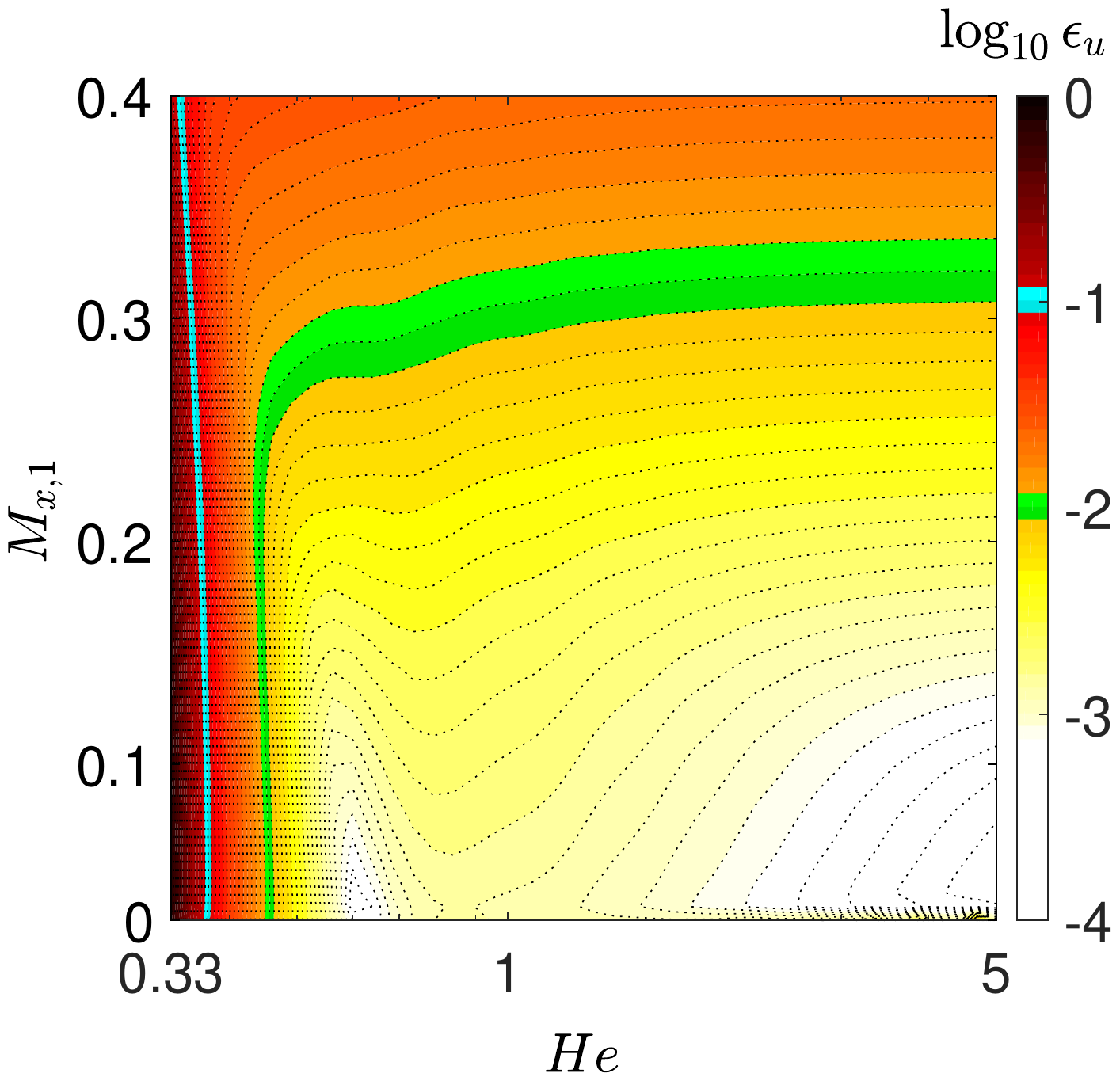}
		\label{Fig:Contour_Mx1_He_u}
	}
	\put (-205,170) {\normalsize$\displaystyle(b)$}
	\vspace*{-0pt}
	\caption{Contour maps  of error coefficients  $\epsilon_p$ and $\epsilon_{u}$ between analytical and numerical results as functions of $He$ and $M_{x,1}$.}
	
	\label{Fig:Contour_Mx1_He} 
	\vspace*{00pt}
\end{figure}
%
\begin{figure}[!h]
	\centering
	\subfigure
	{
		\includegraphics[width=7cm]{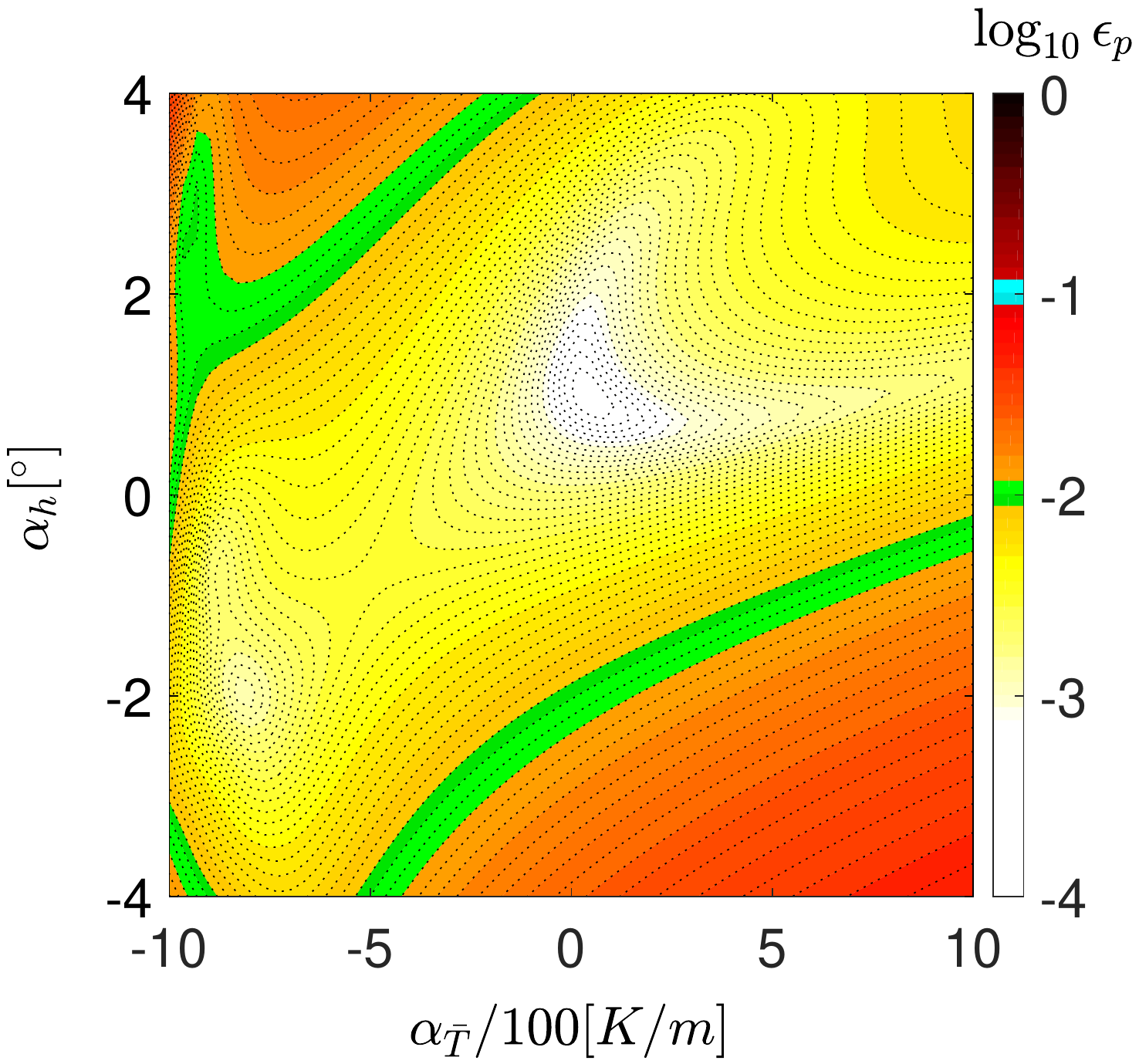}
		\label{Fig:Contour_dh_dTa_p}
	}
	\put (-205,170) {\normalsize$\displaystyle(a)$}
	\vspace*{-0pt}
	\hspace*{30pt}
	\subfigure
	{
		\includegraphics[width=7cm]{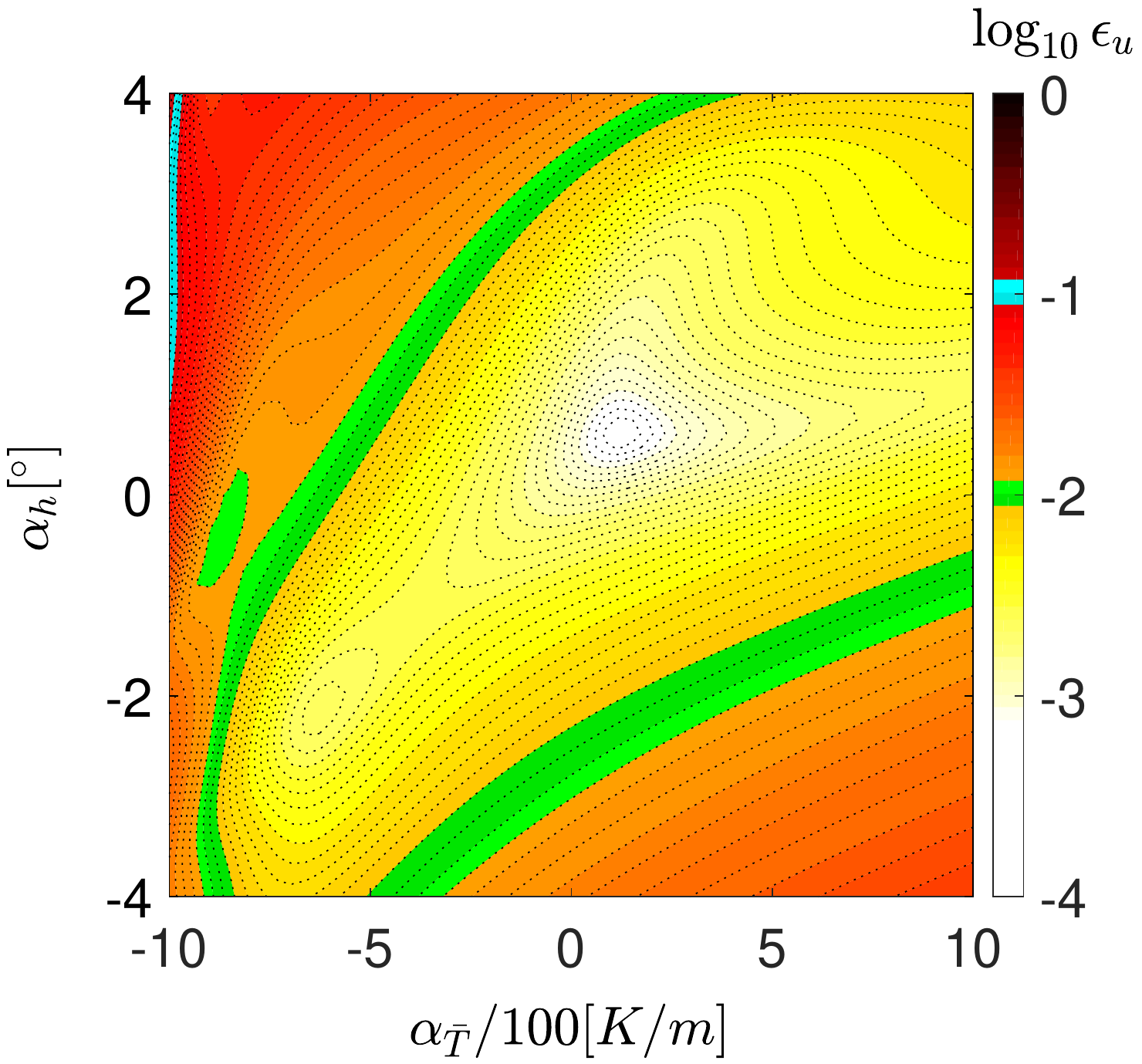}
		\label{Fig:Contour_dh_dTa_u}
	}
	\put (-205,170) {\normalsize$\displaystyle(b)$}
	\vspace*{-0pt}
	\caption{Contour maps of error coefficients $\epsilon_p$ and $\epsilon_{u}$ between analytical and numerical results as functions of $\alpha_h$ and $\alpha_{\overline{T}}$.}
	
	\label{Fig:Contour_dh_dTa} 
	\vspace*{00pt}
\end{figure}
%

In order to further examine the accuracy of the analytical solutions for wide range operating conditions, two error coefficients ($\epsilon_p\left(\mathcal{F}_{p, LEE}, \mathcal{F}_{p, Ana}\right)$ and $\epsilon_{u}\left(\mathcal{F}_{u_x, LEE}, \mathcal{F}_{u_x, Ana}\right)$) are defined,
\begin{equation}
\label{eq:error_coef}
\begin{split}
&\epsilon_{p}\left(\mathcal{F}_{p, LEE}, \mathcal{F}_{p, Ana}\right) = \sqrt{\frac{\sum_{j=1}^{N} \left|\mathcal{F}_{p, LEE}\left(x_{j}\right) - \mathcal{F}_{p, Ana}\left(x_{j}\right) \right|^{2}}{\sum_{j=1}^{N} \left|\mathcal{F}_{p, LEE}\left(x_{j}\right) \right|^{2}}}, \\
&\epsilon_{u}\left(\mathcal{F}_{u_{x}, LEE}, \mathcal{F}_{u_{x}, Ana}\right) = \sqrt{\frac{\sum_{j=1}^{N} \left|\mathcal{F}_{u_{x}, LEE}\left(x_{j}\right) - \mathcal{F}_{u_{x}, Ana}\left(x_{j}\right) \right|^{2}}{\sum_{j=1}^{N} \left|\mathcal{F}_{u_{x}, LEE}\left(x_{j}\right) \right|^{2}}}, \\
&x_{j} = x_{1}, x_{2}, \dots , x_{N}.
\end{split}
\end{equation}
%


Fig.~\ref{Fig:Contour_Mx1_He} shows the error coefficients as functions of inlet Mach number $M_{x,1}$ and Helmholtz number $He$.
The minimum value of Helmholtz number equals the maximum of cut-off frequencies ($He = 0.35$), and  the minimum value of the inlet Mach numbers is set to $M_{x,1} = 0.03$ to avoid the numerical error in the calculation (with decreasing $M_{x,1}$, the duct needs to be segmented to more points in order to reconstruct $\hat{u}_\theta$, see e.g. Fig.~\ref{Fig:n_0_1_2}). 
From Fig.~\ref{Fig:Contour_Mx1_He}, error coefficients are most likely smaller than $0.1$ (indicated by cyan ribbon) and there comes to the conclusion that the analytical solutions perform well for high frequencies and low inlet Mach numbers. Acoustics waves with frequencies lower than cut-off frequency $\omega_{c}$ suffer exponential attenuation and solutions of acoustic field with large Mach numbers lead to unignorable errors.

Fig.~\ref{Fig:Contour_dh_dTa} shows the contour maps of error coefficients vary with non-uniform cross-sectional surface area and mean temperature, which are evaluated  by $\alpha_{h}$ and $\alpha_{\overline{T}}$.
Small changes of chamber gap height are considered and    $\alpha_{h}$ is considered within the range of $\left[ -5^{\circ}, 10^{\circ} \right]$.
 All the parameters not mentioned are set according to Table~\ref{table_parameters}. It is obvious that smaller variations in thickness $h$ and temperature $\overline{T}$ bring higher accuracy; these further validate  that the analytical solutions are suitable for most possible conditions.

\begin{figure}[!h]
	\centering
	\subfigure
	{
		\includegraphics[width=7cm]{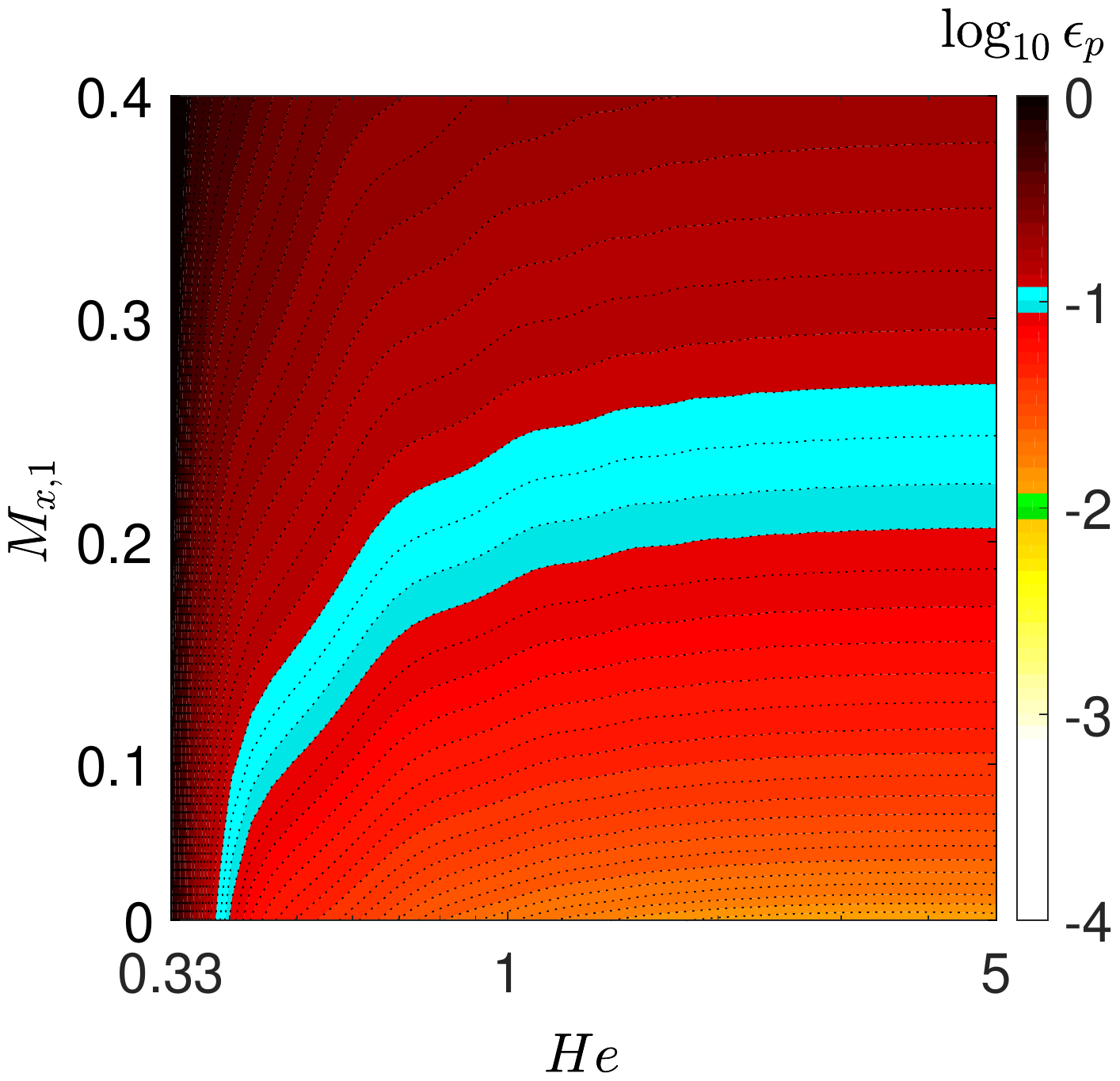}
		\label{Fig:Contour_L_Mx1_He_p}
	}
	\put (-205,170) {\normalsize$\displaystyle(a)$}
	\vspace*{-0pt}
	\hspace*{30pt}
	\subfigure
	{
		\includegraphics[width=7cm]{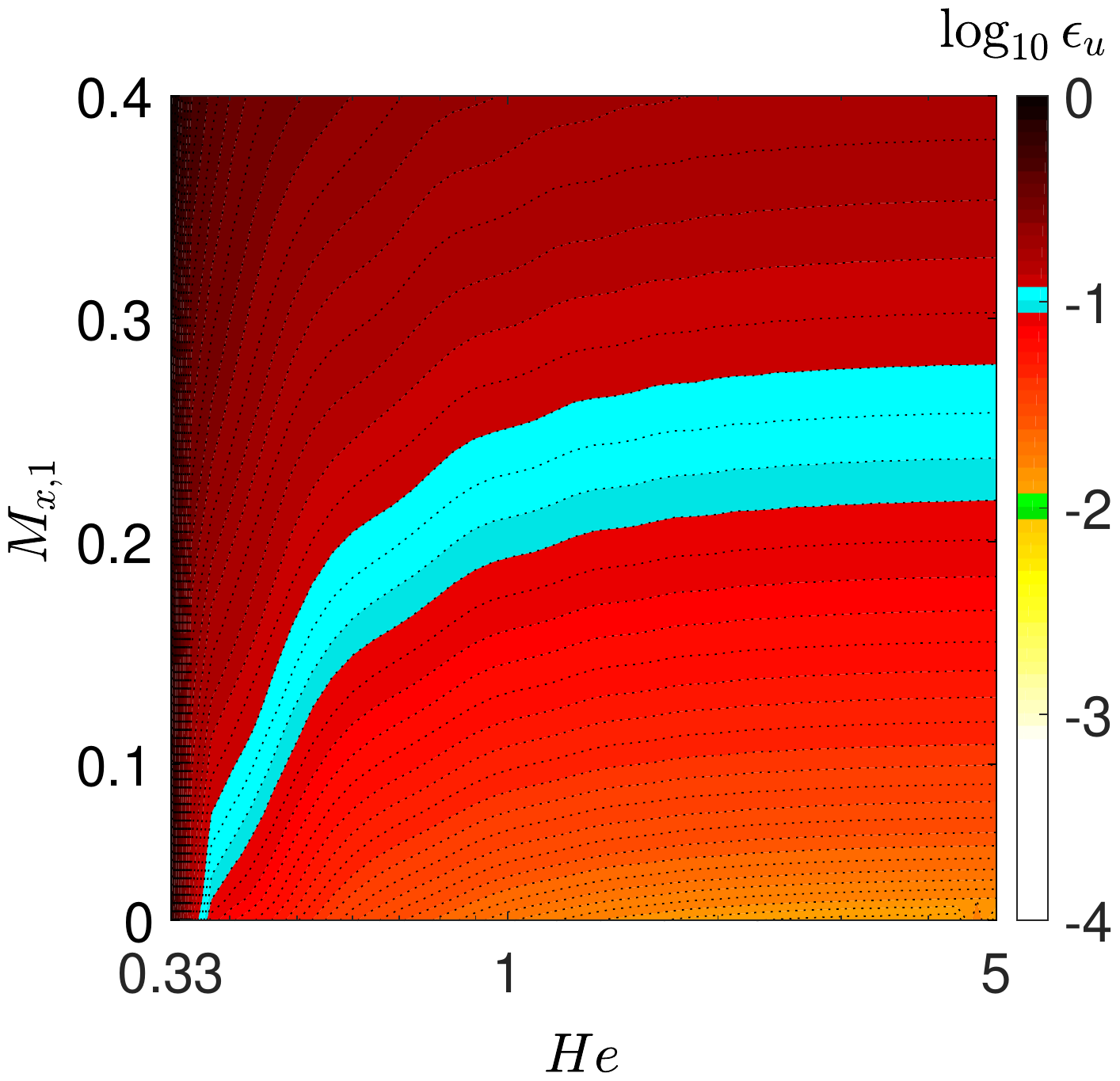}
		\label{Fig:Contour_L_Mx1_He_u}
	}
	\put (-205,170) {\normalsize$\displaystyle(b)$}
	\vspace*{-0pt}
	\caption{Contour maps  of error coefficients $\epsilon_p$ and $\epsilon_{u}$ between simplified analytical and numerical results as functions of $He$ and $M_{x,1}$.}
	
	\label{Fig:Contour_L_Mx1_He} 
	\vspace*{00pt}
\end{figure}
%
\begin{figure}[!h]
	\centering
	\subfigure
	{
		\includegraphics[width=7cm]{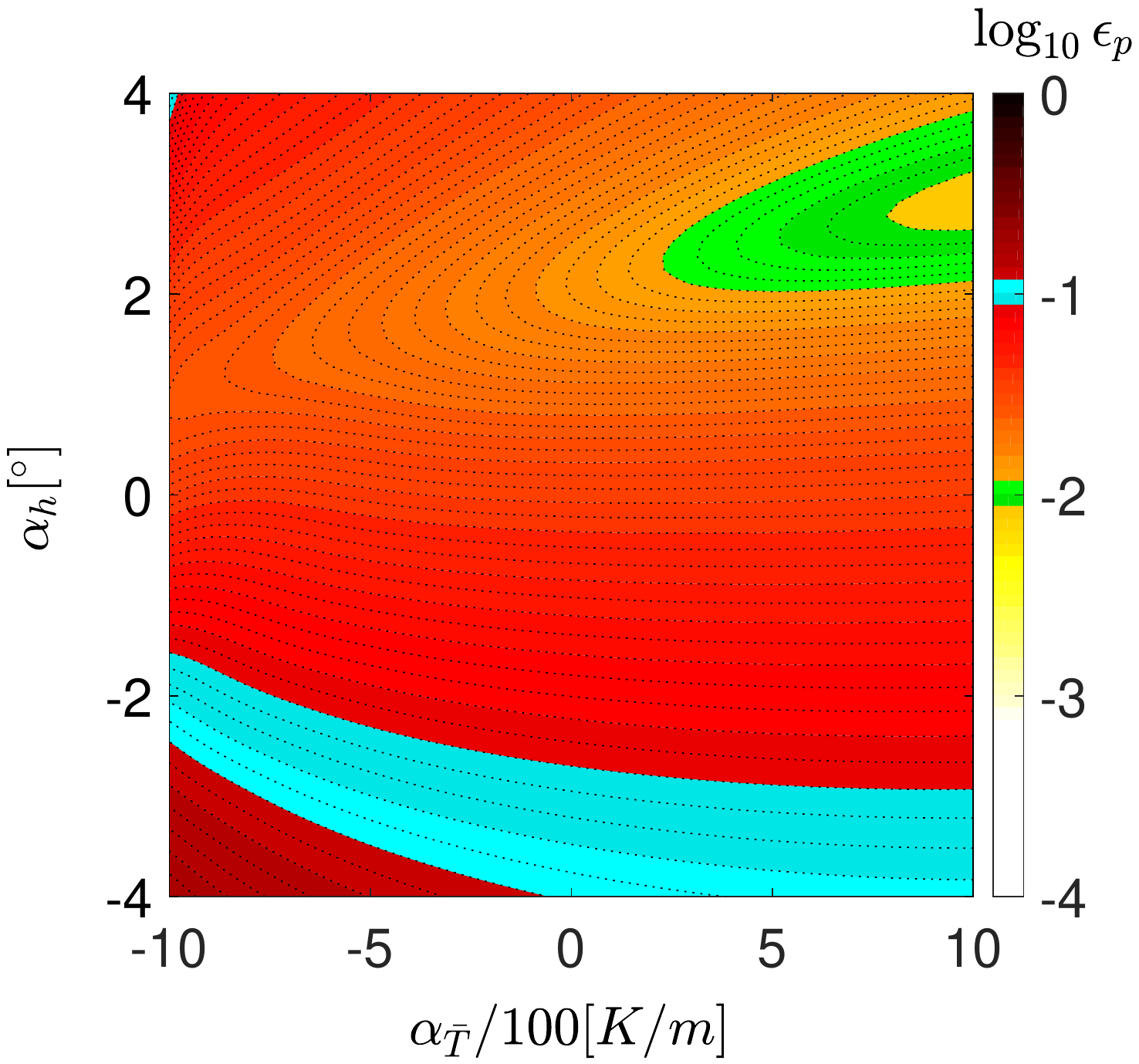}
		\label{Fig:Contour_L_dh_dTa_p}
	}
	\put (-205,170) {\normalsize$\displaystyle(a)$}
	\vspace*{-0pt}
	\hspace*{30pt}
	\subfigure
	{
		\includegraphics[width=7cm]{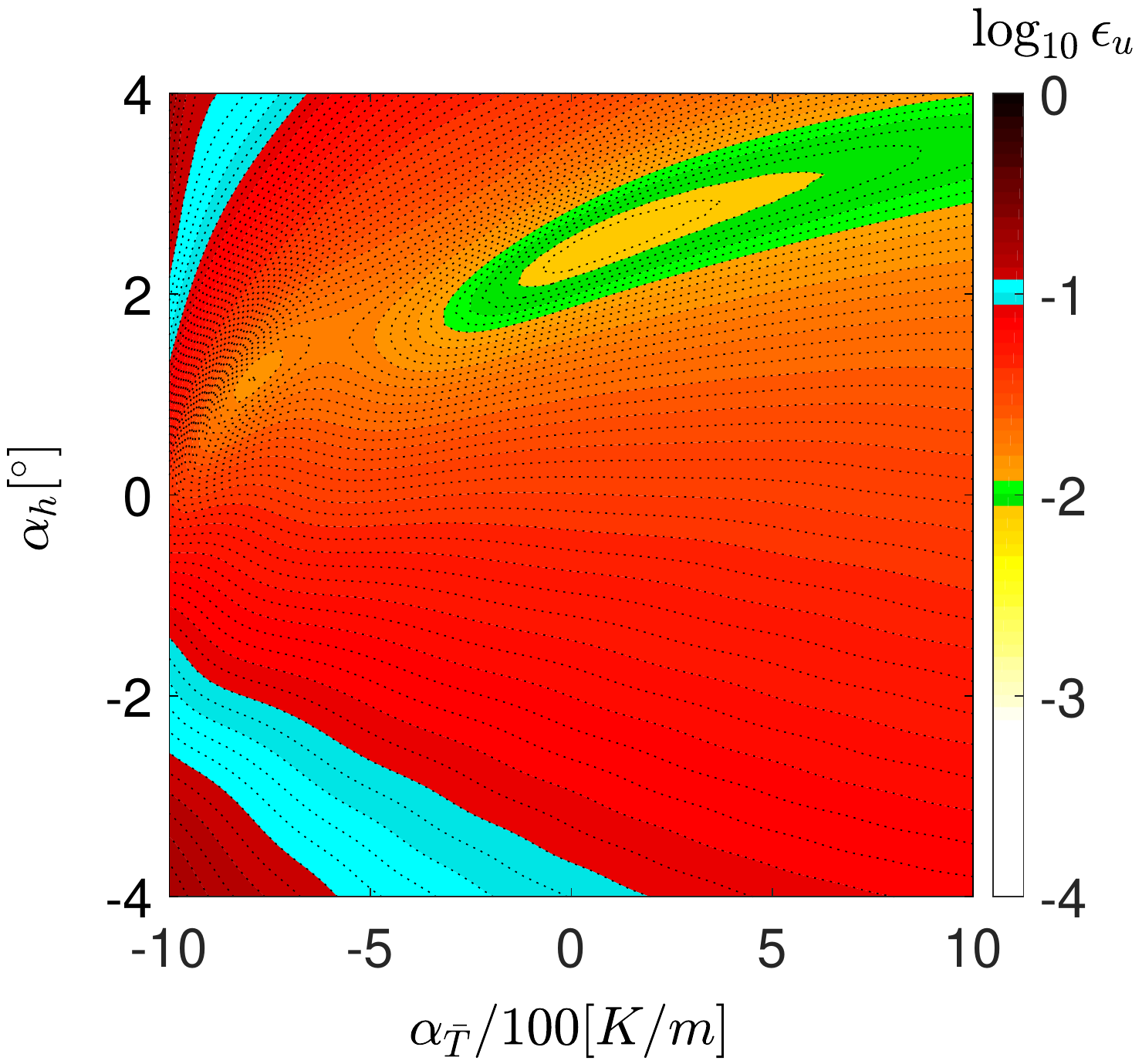}
		\label{Fig:Contour_L_dh_dTa_u}
	}
	\put (-205,170) {\normalsize$\displaystyle(b)$}
	\vspace*{-0pt}
	\caption{Contour maps of error coefficients $\epsilon_p$ and $\epsilon_{u}$ between simplified analytical and numerical results as functions of $\alpha_h$ and $\alpha_{\overline{T}}$.}
	
	\label{Fig:Contour_L_dh_dTa} 
	\vspace*{00pt}
\end{figure}
%

Error coefficients of simplified analytical solutions are also calculated. As shown in Fig.~\ref{Fig:Contour_L_Mx1_He}, errors remain small for Mach numbers lower than 0.2 and are insensitive to the  Helmholtz number $He$. The accuracy is still acceptable. Figure~\ref{Fig:Contour_L_dh_dTa} shows the contour maps of error coefficients as functions of  $\alpha_{h}$ and $\alpha_{\overline{T}}$. The area with highest accuracy is not at the original point, which is more obviously than that in Fig.~\ref{Fig:Contour_dh_dTa}. The errors come from terms dropped in derivation of gap height and temperature gradient and simplified solutions ignore more terms of   $\alpha_{h}$ and $\alpha_{\overline{T}}$.

\subsection{Effects of entropy waves}
\label{subsec:3:3}

\begin{figure}[!h]
	\centering
	\subfigure
	{
		\includegraphics[width=7cm]{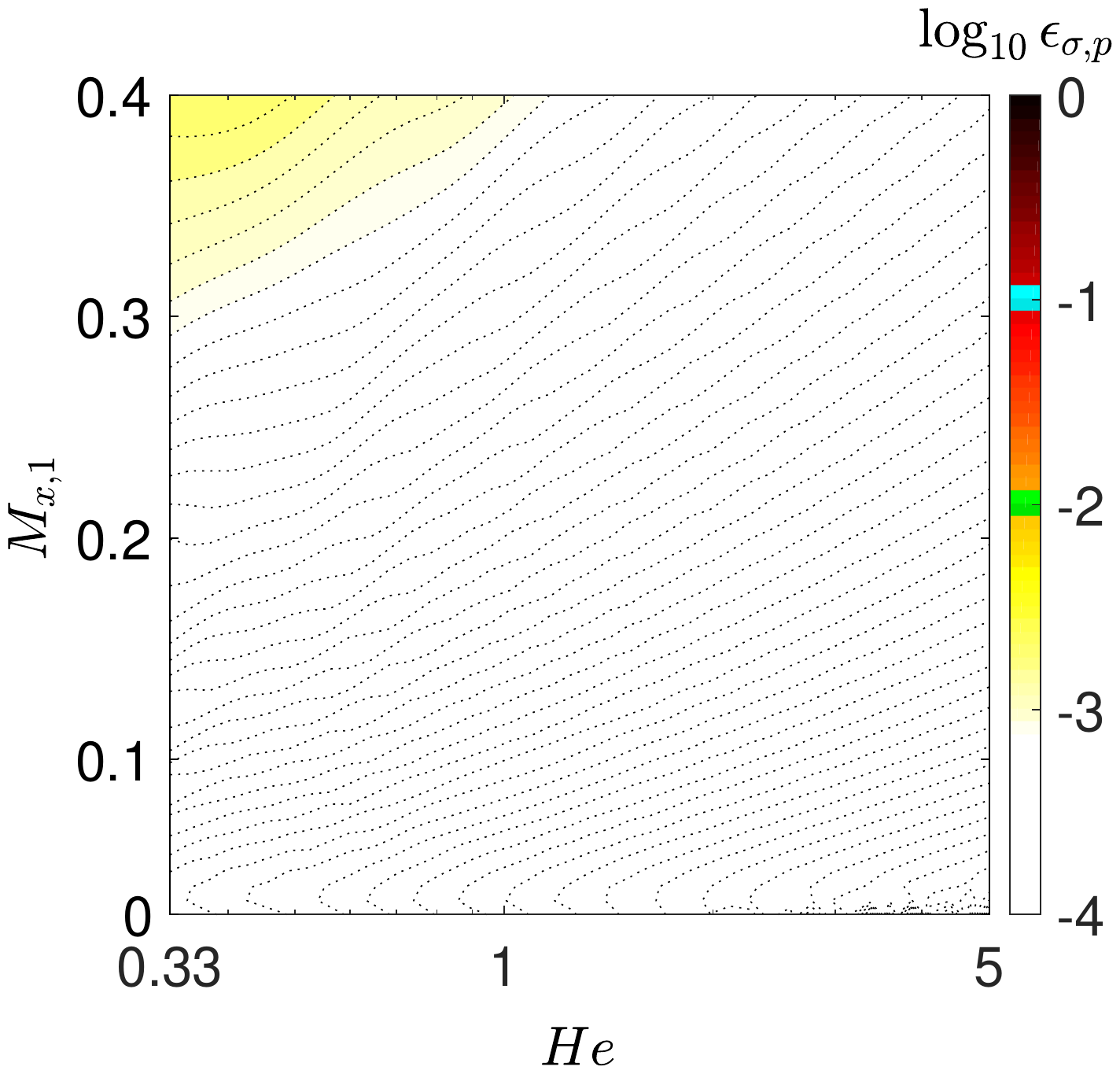}
		\label{Fig:Contour_Sigma_Mx1_He_p}
	}
	\put (-205,170) {\normalsize$\displaystyle(a)$}
	\vspace*{-0pt}
	\hspace*{30pt}
	\subfigure
	{
		\includegraphics[width=7cm]{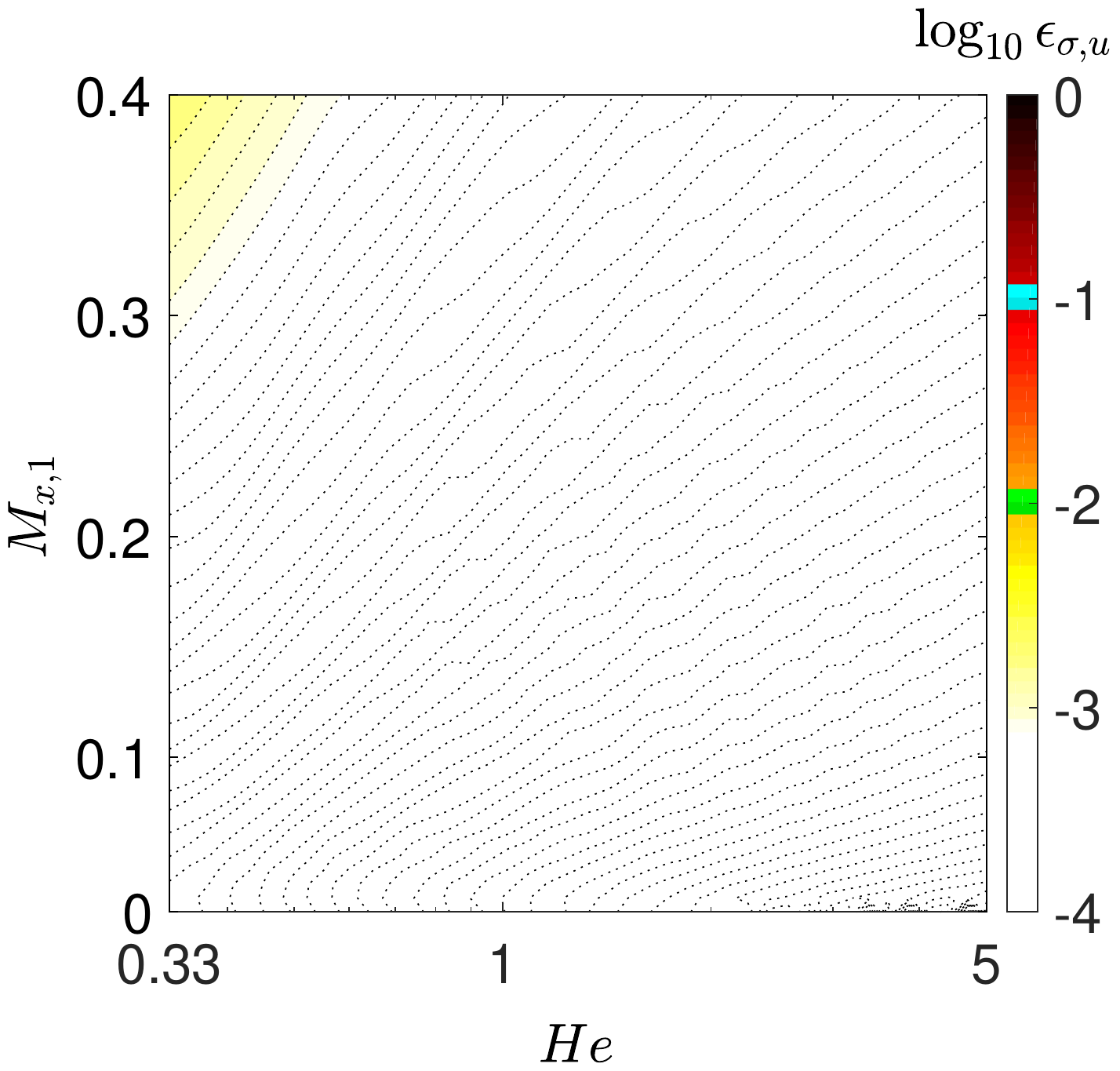}
		\label{Fig:Contour_Sigma_Mx1_He_u}
	}
	\put (-205,170) {\normalsize$\displaystyle(b)$}
	\vspace*{-0pt}
	\caption{Contour maps of error coefficients $\epsilon_{\sigma,p}$ and $\epsilon_{\sigma,u}$ between two numerical results as functions of $M_{x,1}$ and $He$.}
	
	\label{Fig:Contour_Sigma_Mx1_He} 
	\vspace*{00pt}
\end{figure}
%
\begin{figure}[!h]
	\centering
	\subfigure
	{
		\includegraphics[width=7cm]{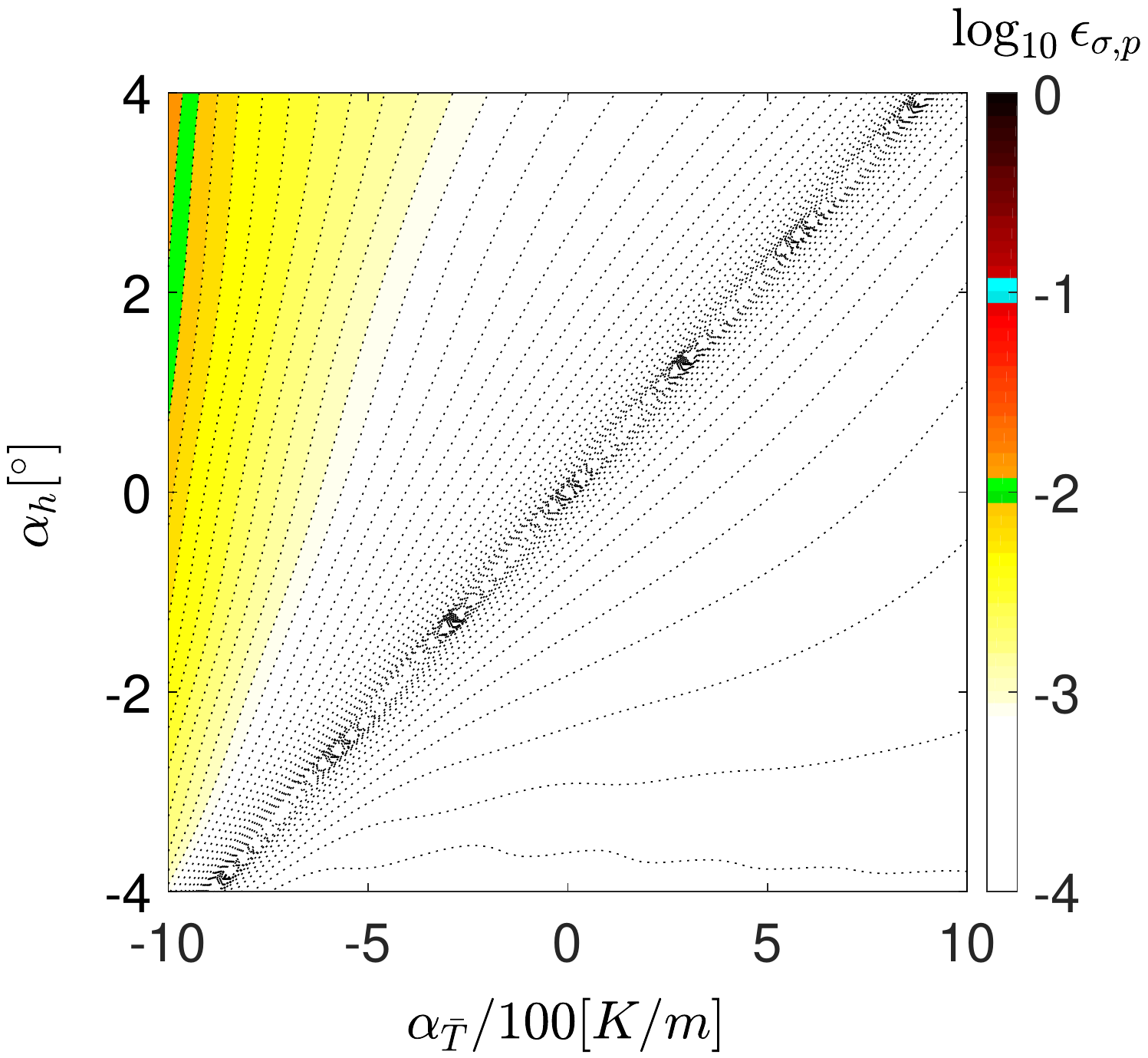}
		\label{Fig:Contour_Sigma_dh_dTa_p}
	}
	\put (-205,170) {\normalsize$\displaystyle(a)$}
	\vspace*{-0pt}
	\hspace*{30pt}
	\subfigure
	{
		\includegraphics[width=7cm]{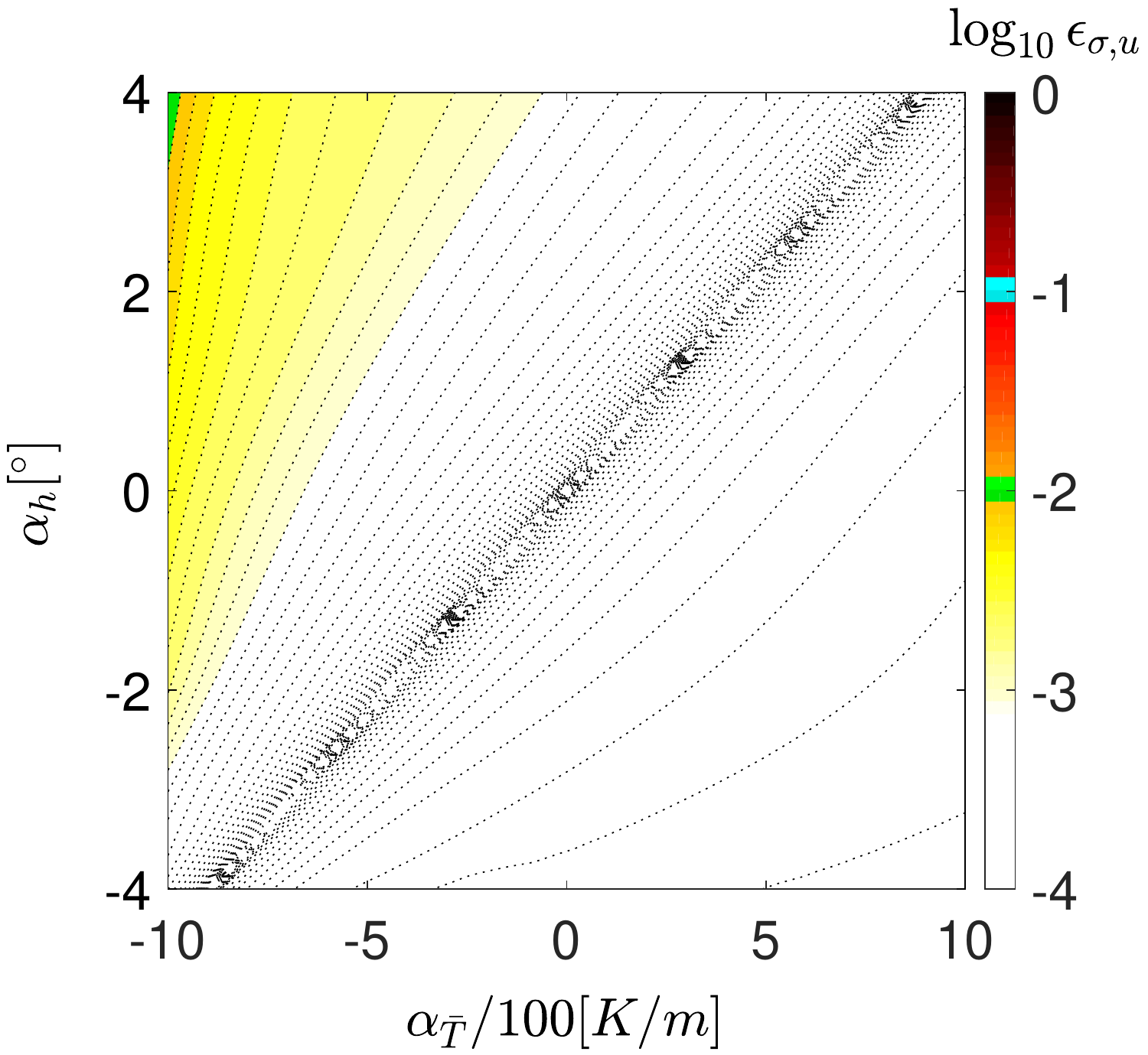}
		\label{Fig:Contour_Sigma_dh_dTa_u}
	}
	\put (-205,170) {\normalsize$\displaystyle(b)$}
	\vspace*{-0pt}
	\caption{Contour maps of error coefficients $\epsilon_{\sigma,p}$ and $\epsilon_{\sigma,u}$ between two numerical results as functions of $\alpha_h$ and $\alpha_{\overline{T}}$. }
	
	\label{Fig:Contour_Sigma_dh_dTa} 
	\vspace*{00pt}
\end{figure}
%
It should be noted that the derivation of analytical solutions and validations are conducted by neglecting the entropy waves. In this section, the effects of entropy waves are examined. The entropy term is retained and the LEEs containing four equations are established (see Appendix \ref{sec:Appendix_LEEs_0_4}).
Following those in Eq.~\eqref{eq:error_coef},  two  error coefficients  $\epsilon_{\sigma,p}$ and $\epsilon_{\sigma,u}$  representing the differences between the predictions from the three equations LEEs the four equations LEEs are defined to examine the effect of entropy waves on the acoustic waves for different operating conditions.

Figure~\ref{Fig:Contour_Sigma_Mx1_He} shows the differences between two numerical methods for the same operating conditions in  Fig.~\ref{Fig:Contour_Mx1_He}. 
The effect of entropy waves increases  with increasing the inlet Mach number $M_{x,1}$ and decreasing the reduced frequency  $He$.  
 Figure~\ref{Fig:Contour_Sigma_dh_dTa} shows that the differences between two numerical methods for the same operating conditions in  Fig.~\ref{Fig:Contour_L_dh_dTa}.
The effect of entropy waves cannot be neglected for relatively large  values of $\alpha_h$ and small values of $\alpha_{\bar{T}}$.  However, the differences remain small and entropy waves can be neglected during the derivations of the analytical solutions  when the assumptions made in Eq.~\eqref{eq:assumps} are satisfied. 
 The analytical method performs well in a wide range of frequencies, inlet Mach numbers, cross-sectional surface area and mean temperature gradients.

\subsection{Results in ducts with arbitrary profiles}
\label{subsec:3:5}

\begin{figure}[!h]
\centering
	\includegraphics[height=5cm]{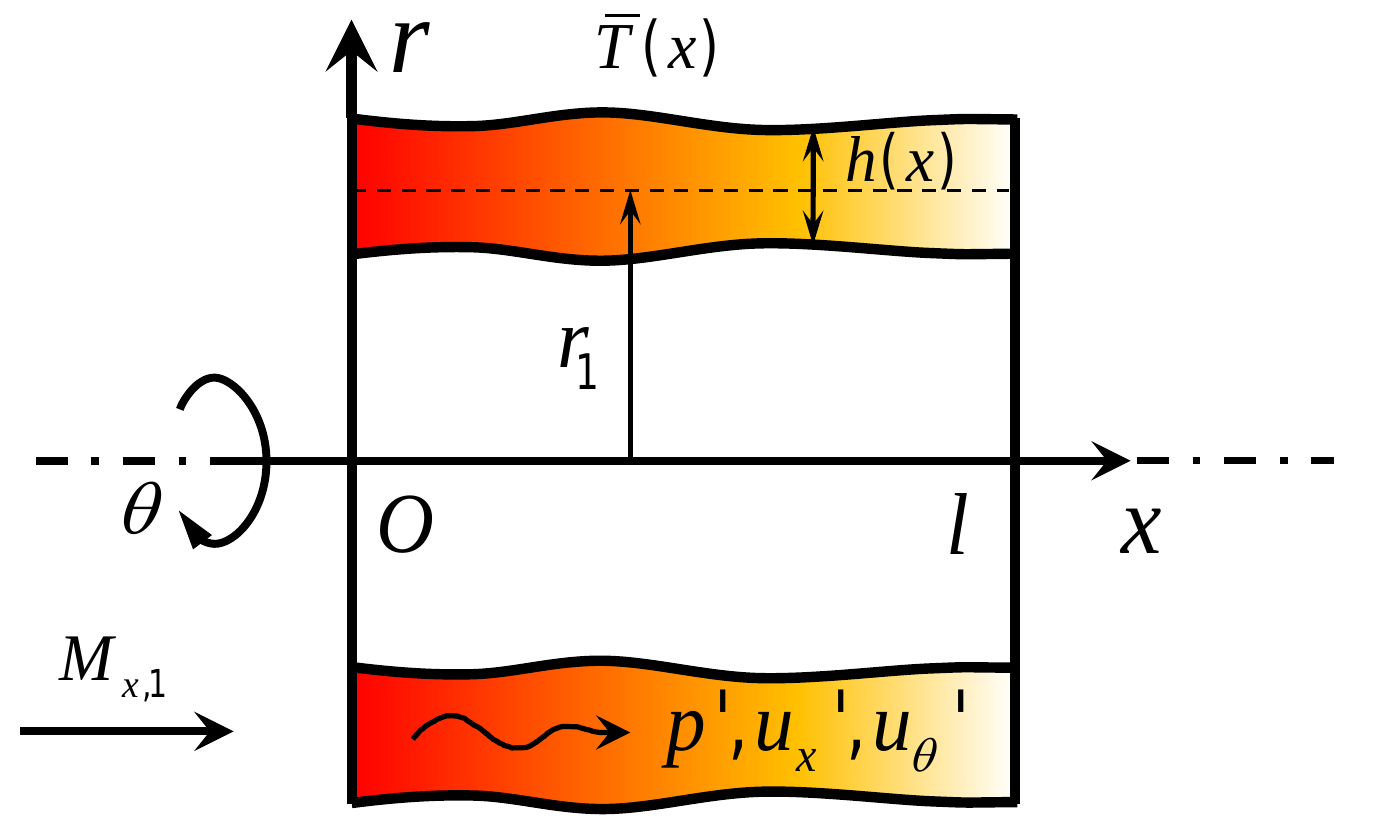}
	\caption{Sketch of a non-uniform cross-sectional surface area annular combustion chamber with mean temperature gradients and mean flow.}
	\label{fig:sketch2}
\end{figure}

\begin{figure}[!h]
	\centering
	\subfigure
	{
		\includegraphics[width=6cm]{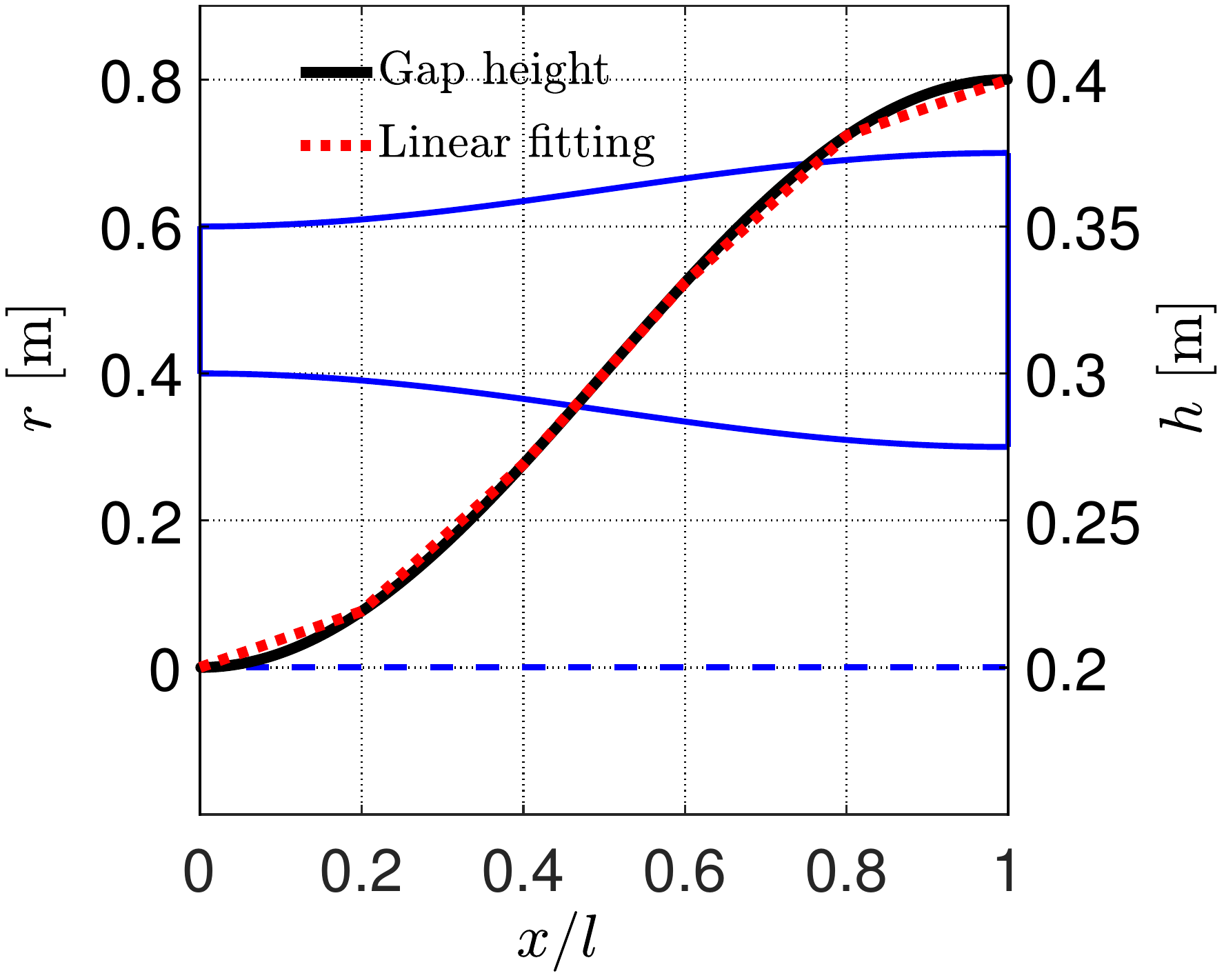}
		\label{Fig:Plot_pw_h}
	}
	\put (-180,130) {\normalsize$\displaystyle(a)$}
	\vspace*{-0pt}
	\hspace*{10pt}
	\subfigure
	{
		\includegraphics[width=5cm]{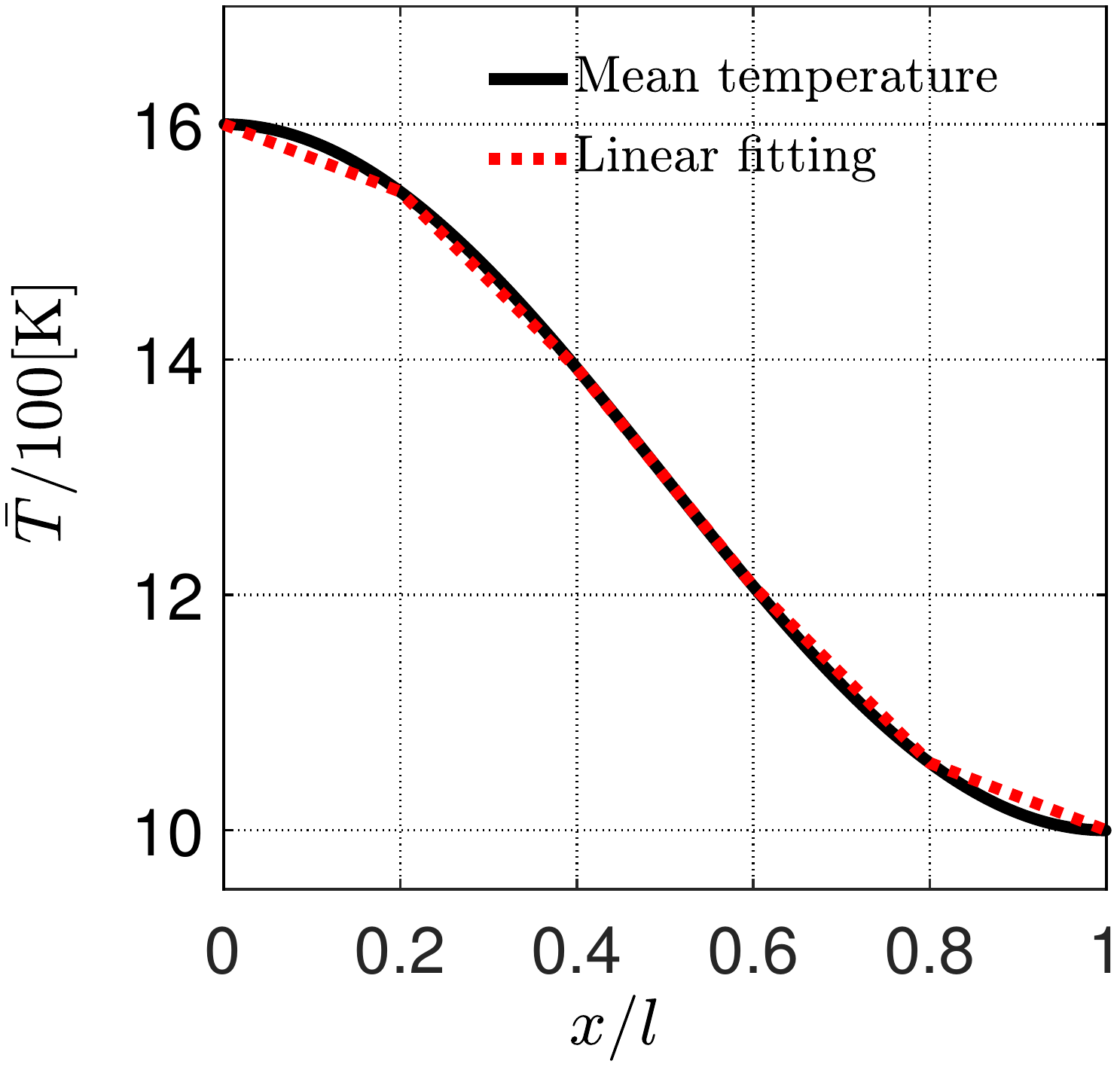}
		\label{Fig:Plot_pw_Ta}
	}
	\put (-155,130) {\normalsize$\displaystyle(b)$}
	\vspace*{-0pt}
	\caption{The gap height and mean temperature gradient of a duct with an arbitrary profile. Thin lines in the left figure (a) represent the shape of the duct and the dashed one stands for the symmetry axis. $N_{s} = 5$.}
	
	\label{Fig:pw_h_Ta} 
	\vspace*{00pt}
\end{figure}

As stated in Ref.~\citep{Li_JSV_2017}, analytical solutions of linear distributions  can be used in a piecewise linear function (PLF) manner for  more complicate profiles (e.g., as sketched in Fig.~\ref{fig:sketch2}). The axial length  is segmented to $N_s$ equal lengths, and  linear profiles of  mean temperature  and   thickness   assumptions are made for each segment. Linear least-squares fittings to the temperature  and thickness profiles are conducted within each segment to yield two PLFs.  The analytical solutions for linear  profiles are then applied to each segment. To examine the performance of the PLF method, one now considers a thin annular duct with following non-uniform thickness and mean temperature, 
\begin{equation}
\label{eq:pw_h_Ta}
\begin{split}
h(x) &= \frac{h_{1}-h_{3}}{2} \cos \left(\pi \frac{x}{l} \right)+\frac{h_{1}+h_{3}}{2},\\
\overline{T}(x) &= \frac{\overline{T}_{1}-\overline{T}_{2}}{2} \cos \left(\pi \frac{x}{l}  \right)+\frac{\overline{T}_{1}+\overline{T}_{2}}{2},
\end{split}
\end{equation}
where $h_{3} = 0.1$ m. Figure~\ref{Fig:pw_h_Ta} shows the duct shape and mean temperature curves, which are both separated into $N_{s} = 5$ segments in the following discussion. The duct has a non-reflective inlet and an open outlet. For analytical solutions, parameters, e.g.,  $\alpha$ and $\beta$,  are calculated independently in each segment and continuous relations are used between two neighbouring segments to obtain arbitrary coefficients $\mathcal{C}_j^+$ and $\mathcal{C}_j^-$, where $j = 1, 2, \dots, N_{s}$. Numerical results are still calculated via the $4000$ points uniform grid. 

%

\begin{figure}[!h]
	\hspace*{13pt}
	\subfigure
	{
		\includegraphics[height=7cm]{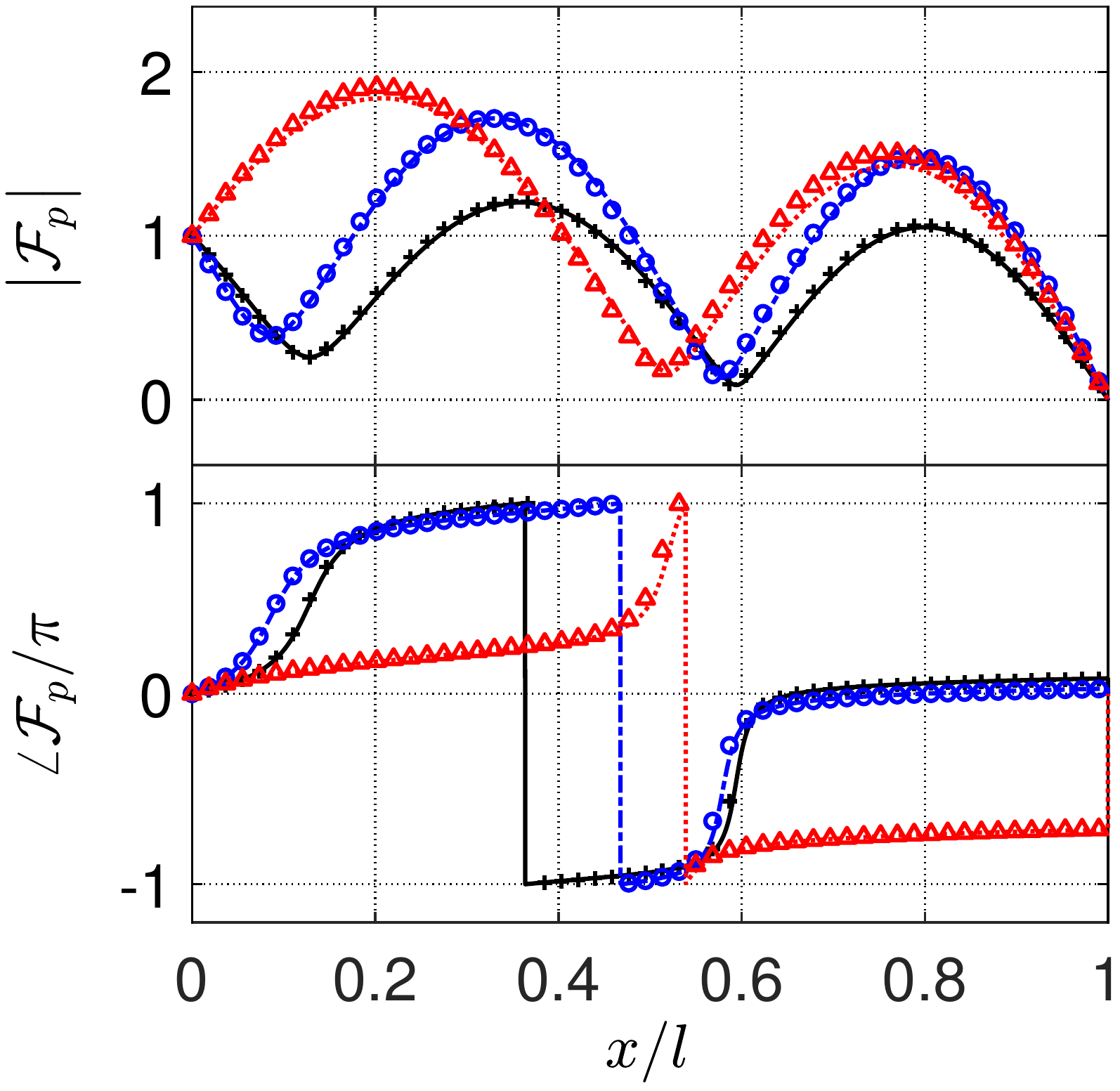}
		\label{Fig:pw_n3_p}
	}
	\put (-200,180) {\normalsize$\displaystyle(a)$}
	\vspace*{-0pt}
	\hspace*{10pt}
	\subfigure
	{
		\includegraphics[height=7cm]{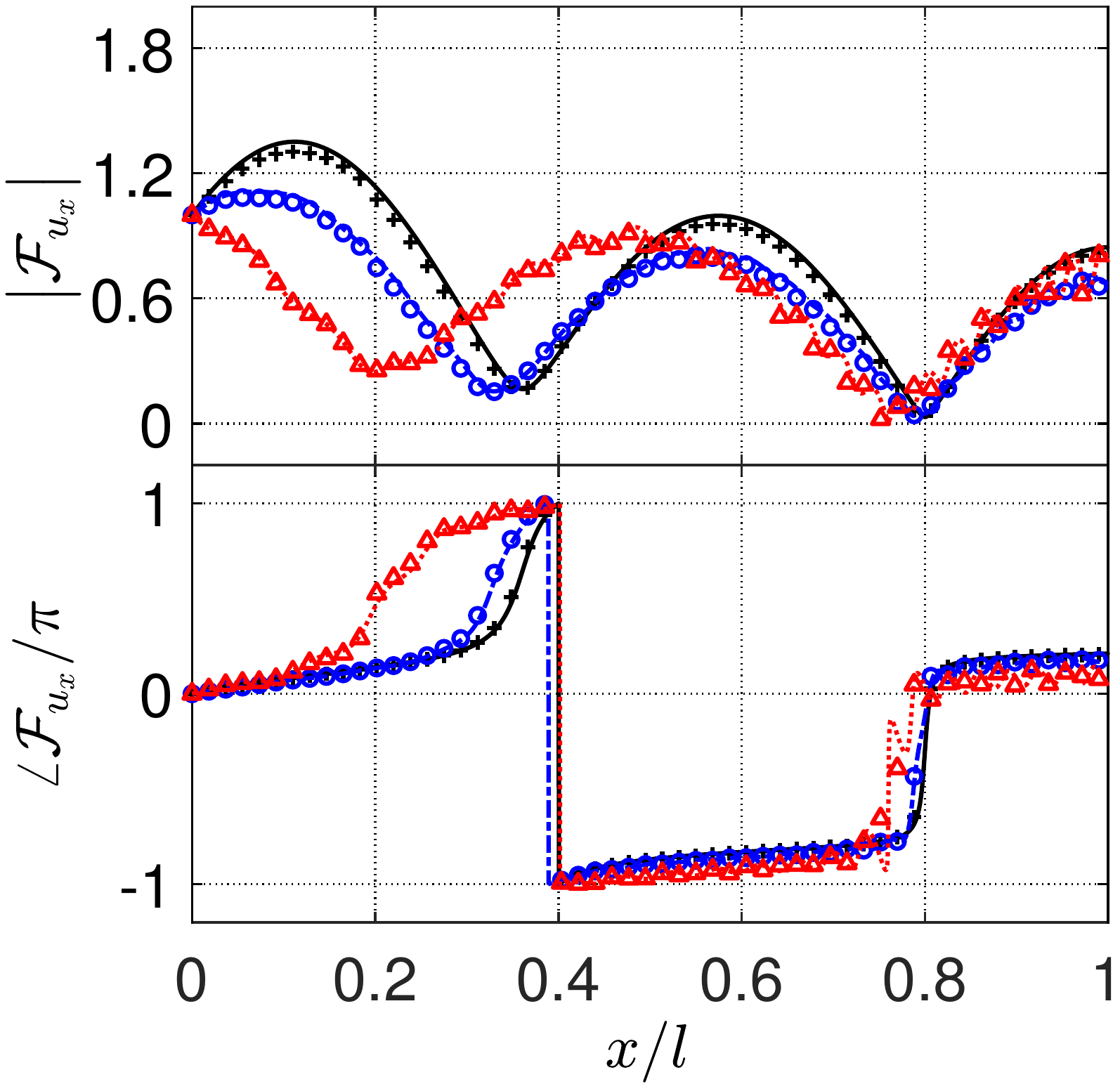}
		\label{Fig:pw_n3_ux}
	}
	\put (-200,180) {\normalsize$\displaystyle(b)$}
	\vspace*{-0pt}\\
	\hspace*{13pt}
	\subfigure
	{
		\includegraphics[height=7cm]{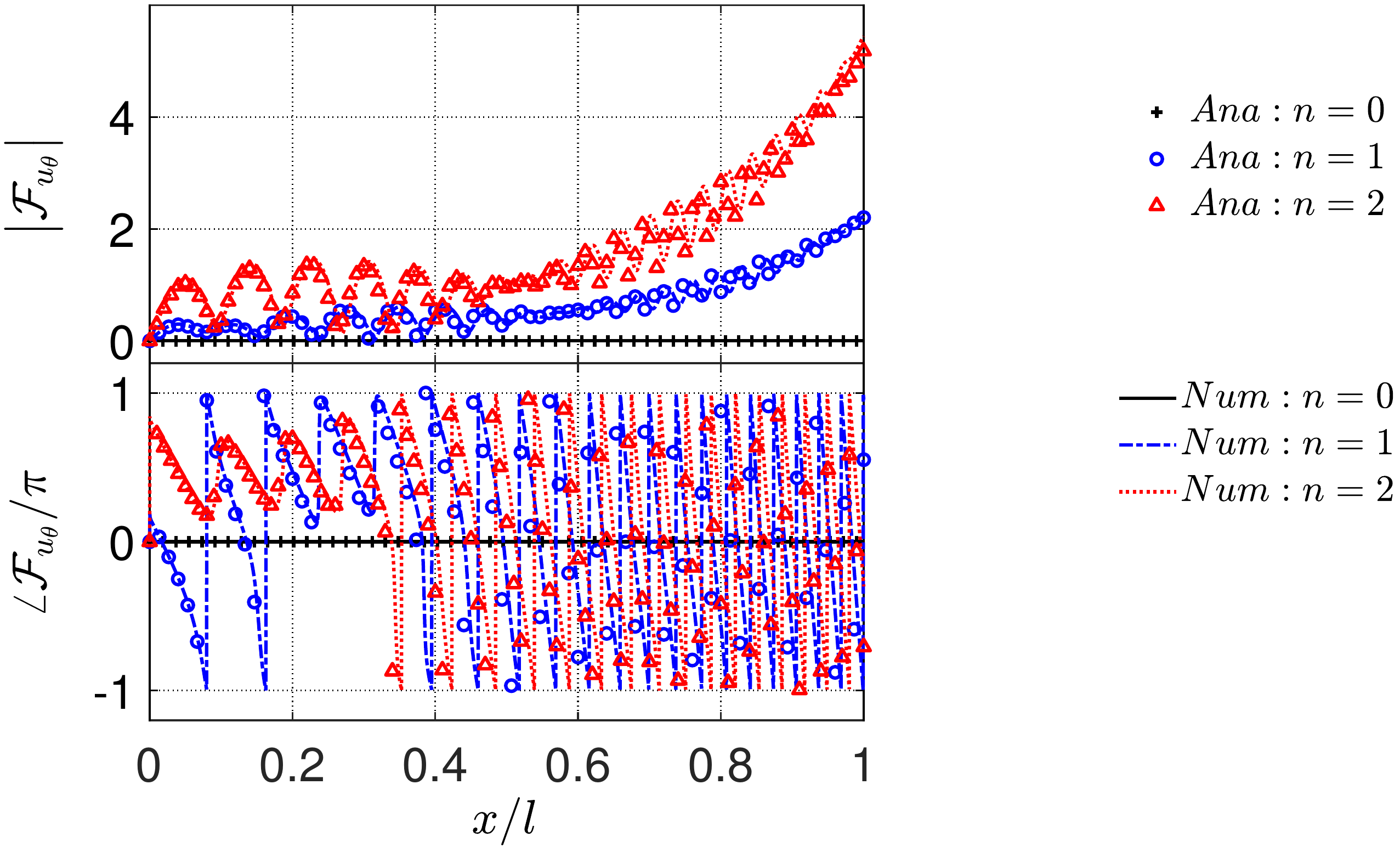}
		\put (-323,180) {\normalsize$\displaystyle(c)$}
		\label{Fig:pw_n3_uth}
	}
	\vspace*{-0pt}
	\caption{Axial distributions of $\mathcal{F}_p$, $\mathcal{F}_{u_x}$ and $\mathcal{F}_{u_\theta}$ calculated by the analytical solution used in the PLF manner ($Ana$) and numerical ($Num$) methods for  different circumferential wavenumbers ($n = 0, 1, 2$). The thickness and mean temperature gradient are as shown in Fig.~\ref{Fig:pw_h_Ta} and the remaining parameters are shown in Table~\ref{table_parameters}.}
	
	\label{Fig:pw_n_0_1_2} 
	\vspace*{00pt}
\end{figure}
%

Figure~\ref{Fig:pw_n_0_1_2} shows the transfer functions of the acoustic field predicted by the analytical solutions and three equations LEEs method for three circumferential wavenumbers $n = 0, 1$ and  2. Good agreements are again found, indicating that the analytical solutions applied in the PLF manner are suitable for the thin annular combustion chamber with  arbitrary profiles of mean temperature and thickness. 


\section{Conclusions}
\label{sec:5}

This article provides an analytical method to obtain the solutions of acoustic field in thin annular combustion chambers with non-uniform cross-sectional surface area and mean temperature gradients. Under the assumptions of high frequency, low Mach number, small growth rate and small non-uniformity (Eq.~\eqref{eq:assumps}), a wave equation for the pressure perturbation (Eq.~\eqref{eq:wave_equation}) is derived and then solved by a modified WKB method approximately. The solutions (Eqs.~\eqref{eq:wave_hat_p}-\eqref{eq:ux_D}) perform well across a wide range of parameters conditions and can reconstruct the acoustic field in  thin annular ducts with arbitrary cross-sectional surface area variations and mean temperature gradients by the PLF method. Errors from the derivation (Eq.~\eqref{eq:momentum_x_LEEs}, \eqref{eq:assumps} and \eqref{eq:dMx}) are checked carefully to confirm the accuracy. The simplified analytical solutions (Eqs.~\eqref{eq:P+-_L}-\eqref{eq:ux_D_L}) also have acceptable accuracy under the summarised  assumptions as in Eq.~\eqref{eq:assumps}.

The solutions provide an explicit method to get the axial distribution of acoustic perturbations in annular chambers, which can be used to predict the thermoacoustic instabilities through the low-order network models. Different from previous researches \citep{Li_JSV_2017,Li_JSV_2020,Li_JSV_2021}, frequencies and axial wave numbers are regarded as complex numbers, which allows the prediction of resonance frequencies as well as their growth rates. This is helpful in reducing the cost of calculation and analysis of physical mechanism of real annular combustion chambers.

\section*{Acknowledgement}
The authors would like to gratefully acknowledge financial support from the Chinese National Natural Science Funds for National Natural Science Foundation of China (Grant no. 11927802), and National Major Science and Technology Projects of China (2017-III-0004-0028). The European Research Council grant AFIRMATIVE (2018–2023, Grant Number 772080) is also gratefully acknowledged.  
%

%
\appendix
%

\section{Relations of time-averaged flow properties}
\label{sec:Appendix_D_param}

The time-averaged parts of Eqs.~\eqref{eq:mass_thin}-\eqref{eq:ideal_gas} are as follows,
\begin{equation}
\label{eq:mean_mass_thin}
\alpha + \frac{1}{\bar{u}_{x}} \frac{\mathrm{d} \bar{u}_{x}}{\mathrm{d} x} + \frac{1}{\bar{\rho}} \frac{\mathrm{d} \bar{\rho}}{\mathrm{d} x} = 0,
\end{equation}
\begin{equation}
\label{eq:mean_x_thin}
\frac{\mathrm{d} \bar{p}}{\mathrm{d} x} + \bar{\rho} \bar{u}_{x} \frac{\mathrm{d} \bar{u}_{x}}{\mathrm{d} x} = 0,
\end{equation}
\begin{equation}
\label{eq:mean_energy_thin}
\frac{\mathrm{d} \bar{p}}{\mathrm{d} x} + \gamma \bar{p}\left(\alpha + \frac{1}{\bar{u}_{x}} \frac{\mathrm{d} \bar{u}_{x}}{\mathrm{d} x}\right) = \left( \gamma - 1 \right) \frac{\bar{\dot{q}}}{\bar{u}_{x}},
\end{equation}
\begin{equation}
\label{eq:mean_ideal_gas}
\bar{p} = \bar{\rho} R_g \overline{T}.
\end{equation}
The time-averaged flow properties are uniform in the circumferential direction because of the  spatial periodicity, which means they are only the functions of axial position $x$. 
Based on the above relations, one can obtain the  normalised differential of time-averaged flow properties  as the functions of $M_x$, $\alpha$ and $\beta$, expressed as:
\begin{equation}
\label{eq:D_param}
\centering
\begin{aligned}
&\frac{1}{\bar{p}} \frac{\mathrm{d} \bar{p}}{\mathrm{d} x} = - \frac{\gamma M_x^{2}}{1- \gamma M_x^{2}} \left(\beta - \alpha \right),\\
&\frac{1}{\bar{u}_{x}} \frac{\mathrm{d} \bar{u}_{x}}{\mathrm{d} x} = \frac{1}{1- \gamma M_x^{2}} \left(\beta - \alpha \right),\\
&\frac{1}{k_{r}} \frac{\mathrm{d} k_{r}}{\mathrm{d} x} = \frac{1}{k_{i}} \frac{\mathrm{d} k_{i}}{\mathrm{d} x} = - \frac{1}{\bar{c}} \frac{\mathrm{d} \bar{c}}{\mathrm{d} x}= - \frac{1}{2} \beta,\\
&\frac{1}{\bar{\rho}} \frac{\mathrm{d} \bar{\rho}}{\mathrm{d} x} = 
- \frac{1}{1- \gamma M_x^{2}} \left(\beta - \gamma M_x^{2} \alpha \right),\\
&\frac{1}{M_x}\frac{\mathrm{d} M_x}{\mathrm{d} x} =  
\frac{1}{2 \left( 1- \gamma M_x^{2} \right)} \left(\left( 1 + \gamma M_x^{2} \right) \beta - 2 \alpha \right).
\end{aligned}
\end{equation}


\section{Four equations  LEEs}
\label{sec:Appendix_LEEs_0_4}

Linearised Euler equations with different forms are mentioned as the accurate results or to derive the analytical solutions.
%
%
%
%
%
%
Numerical results of $4$ LEEs method are calculated by LEEs with entropy waves,
\begin{equation}
\label{eq:mass_LEEs4}
\bar{u}_x^{} \frac{\mathrm{d} \hat{\rho}}{\mathrm{d} x}  +  \left(\mathrm{i} \omega +  \alpha \bar{u}_{x}  +  \frac{\mathrm{d} \bar{u}_{x}}{\mathrm{d} x} \right) \hat{\rho}  +  \bar{\rho} \frac{\mathrm{d} \hat{u}_x{}}{\mathrm{d} x} 
 +  \left( \alpha \bar{\rho} + \frac{\mathrm{d} \bar{\rho}}{\mathrm{d} x} \right)\hat{u}_{x}  +  \frac{\mathrm{i} n}{r} \bar{\rho} \hat{u}_\theta{} = 0,
\end{equation}
\begin{equation}
\label{eq:energy_x_LEEs4}
\bar{u}_x^{} \frac{\mathrm{d} \hat{p}}{\mathrm{d} x}  +  \left(\mathrm{i} \omega +  \gamma \alpha \bar{u}_{x}  +  \gamma \frac{\mathrm{d} \bar{u}_{x}}{\mathrm{d} x} \right) \hat{p}  +  \gamma \bar{p} \frac{\mathrm{d} \hat{u}_x{}}{\mathrm{d} x} 
 +  \left(\gamma \alpha\bar{p} + \frac{\mathrm{d} \bar{p}}{\mathrm{d} x} \right)\hat{u}_{x}    + \gamma \bar{p} \frac{\mathrm{i} n}{r} \hat{u}_\theta{} = 0,
\end{equation}
\begin{equation}
\label{eq:momentum_x_LEEs4}
\frac{1}{\bar{\rho}}\frac{\mathrm{d} \hat{p} }{\mathrm{d} x}
+ \bar{u}_{x}\frac{\mathrm{d} \hat{u}_{x}}{\mathrm{d} x}
+  \left(\mathrm{i} \omega + \frac{\mathrm{d} \bar{u}_{x}}{\mathrm{d} x} \right) \hat{u}_x{} 
+ \frac{\bar{u}_{x}}{\bar{\rho}} \frac{\mathrm{d} \bar{u}_{x}}{\mathrm{d} x} \hat{\rho}= 0,
\end{equation}
\begin{equation}
\label{eq:momentum_theta_LEEs4}
\frac{1}{\bar{\rho}} \frac{\mathrm{i} n}{r}  \hat{p} + \bar{u}_x{} \frac{\mathrm{d} \hat{u}_\theta{} }{\mathrm{d} x}
+ \mathrm{i} \omega \hat{u}_\theta{}=0.
\end{equation}


\section{Comparison to other analytical solutions}
\label{sec:Appendix_com_ana}

\subsection{Solutions of acoustic field in annular ducts with variable thickness and isentropic flow}
\label{sec:Appendix_com_ana_annular}

In our previous work \citep{Li_JSV_2021}, solutions of acoustic field in ducts with variable thickness and average radius are derived. The flow is isentropic and the imaginary part of  the axial wavenumber $k$ is ignored. The solutions in Ref.~\citep{Li_JSV_2021} and solutions in this article can be reduced to the same case: annular duct with only variable thickness and isentropic flow. For Eqs.~\eqref{eq:wave_hat_p}-\eqref{eq:ux_D}, the transformation can be achieved by setting,
\begin{equation}
\label{eq:trans_1}
\begin{split}
\beta = \frac{\gamma - 1}{1-M^{2}_{x}} M^{2}_{x} \alpha, \quad \epsilon = 0.
\end{split}
\end{equation}
The final solutions keep the similar form as Eqs.~\eqref{eq:wave_hat_p}-\eqref{eq:ux_D} and essential expressions of the reduced solutions are as follows, 
\begin{equation}
\label{eq:b_1}
b^\pm_{1} = \frac{M_x \mp \lambda}{1 - M_x^2} k_{r},
\end{equation}
\begin{equation}
\label{eq:a_1}
a^\pm_{1} = -\frac{1}{2} \alpha - \frac{1}{2 k^*}\frac{\mathrm{d} k^*}{\mathrm{d} x} 
\pm  \lambda \alpha M_{x} - \frac{1}{2}\alpha M^{2}_{x},
\end{equation}
\begin{equation}
\label{eq:uxB_1}
\mathcal{B}^\pm_{1} =  \mathrm{i} k^{*} \pm \frac{1}{2} \alpha - \lambda \alpha M_{x} \pm \frac{5 - \gamma}{4} \alpha M^{2}_{x},
\end{equation}
\begin{equation}
\label{eq:uxD_1}
\mathcal{D}_{1} = \mathrm{i} k_{r} - \frac{2 M_{x}}{1 - M^{2}_{x}} \alpha = \mathrm{i} k_{r} -2 \alpha M_{x}.
\end{equation}
Solutions in Ref.~\citep{Li_JSV_2021} have the same expressions as Eqs.~\eqref{eq:b_1}-\eqref{eq:uxD_1} under the assumptions of Eq.~\eqref{eq:assumps}.

\subsection{Solutions of acoustic field in straight ducts with mean temperature gradient and mean flow}
\label{sec:Appendix_com_ana_1D}

Then Eqs.~\eqref{eq:wave_hat_p}-\eqref{eq:ux_D} are reduced to cases of $1D$ straight ducts with only mean temperature gradient and mean flow (Ref.~\citep{Li_JSV_2017}). The transformation conditions are,
\begin{equation}
\label{eq:trans_2}
\begin{split}
\beta = \left( \gamma M^{2}_{x} - 1 \right) \delta, \quad \epsilon = n = \alpha = 0,
\end{split}
\end{equation}
where $\delta = \frac{1}{\bar{\rho}} \frac{\mathrm{d} \bar{\rho}}{\mathrm{d} x}$ and annular ducts become straight ducts when $n = 0$. 
The final solutions keep the similar form as Eqs.~\eqref{eq:wave_hat_p}-\eqref{eq:ux_D} and essential expressions of the reduced solutions are as follows, 
\begin{equation}
\label{eq:b_2}
b^\pm_{2} = \frac{\mp k_{r}}{1 \pm M_x},
\end{equation}
\begin{equation}
\label{eq:a_2}
a^\pm_{2} = \frac{1}{2} \delta - \frac{1}{2 k^*}\frac{\mathrm{d} k^*}{\mathrm{d} x} 
\pm  \frac{\gamma + 1}{2} \delta M_{x} - \frac{\gamma +1}{2} \delta M^{2}_{x}\\
= \frac{1}{4} \delta \pm  \frac{\gamma + 1}{2} \delta M_{x} - \frac{\gamma +2}{4} \delta M^{2}_{x},
\end{equation}
\begin{equation}
\label{eq:uxB_2}
\mathcal{B}^\pm_{2} =  \mathrm{i} k_{r} \mp \frac{1}{4} \delta - \frac{\gamma + 1}{2} \delta M_{x} \mp \frac{3 \gamma - 7}{4} \delta M^{2}_{x},
\end{equation}
\begin{equation}
\label{eq:uxD_2}
\mathcal{D}_{2} = \mathrm{i} k_{r} - \delta M_{x}.
\end{equation}
Though derivations of two solutions are based on different parameters ($\alpha$, $\beta$ and $\delta$),  Eqs.~\eqref{eq:wave_hat_p}-\eqref{eq:ux_D} accurately coincide with the solutions in Ref.~\citep{Li_JSV_2017}.

%
%
%
%
%
%
%
%
\bibliographystyle{elsarticle-num}
\bibliography{AST_2022_references}

\begin{thebibliography}{10}
\expandafter\ifx\csname url\endcsname\relax
  \def\url#1{\texttt{#1}}\fi
\expandafter\ifx\csname urlprefix\endcsname\relax\def\urlprefix{URL }\fi
\expandafter\ifx\csname href\endcsname\relax
  \def\href#1#2{#2} \def\path#1{#1}\fi

\bibitem{Ruan_AST_2020}
C.~Ruan, F.~Chen, T.~Yu, W.~Cai, X.~Li, X.~Lu, Experimental study on flame/flow
  dynamics in a multi-nozzle gas turbine model combustor under
  thermo-acoustically unstable condition with different swirler configurations,
  Aerosp. Sci. Technol. 98 (2020) 105692.
\newblock \href {https://doi.org/10.1016/j.ast.2020.105692}
  {\path{doi:10.1016/j.ast.2020.105692}}.

\bibitem{Chen_AST_2019}
F.~Chen, C.~Ruan, T.~Yu, W.~Cai, Y.~Mao, X.~Lu, Effects of fuel variation and
  inlet air temperature on combustion stability in a gas turbine model
  combustor, Aerosp. Sci. Technol. 92 (2019) 126--138.
\newblock \href {https://doi.org/10.1016/j.ast.2019.05.052}
  {\path{doi:10.1016/j.ast.2019.05.052}}.

\bibitem{Lilei_AST_2015}
L.~Li, D.~Zhao, Prediction of stability behaviors of longitudinal and
  circumferential eigenmodes in a choked thermoacoustic combustor, Aerosp. Sci.
  Technol. 46 (2015) 12--21.
\newblock \href {https://doi.org/10.1016/j.ast.2015.06.024}
  {\path{doi:10.1016/j.ast.2015.06.024}}.

\bibitem{LiLei_AST_2018}
L.~Li, D.~Zhao, X.~Sun, Nonorthogonality analysis of acoustics and vorticity
  modes: Should thermoacoustic energy norm be time-invariant?, Aerosp. Sci.
  Technol. 77 (2018) 149--155.
\newblock \href {https://doi.org/10.1016/j.ast.2018.02.036}
  {\path{doi:10.1016/j.ast.2018.02.036}}.

\bibitem{Si_AST_2013}
H.~Si, W.~Shen, W.~Zhu, Effect of non-uniform mean flow field on acoustic
  propagation problems in computational aeroacoustics, Aerosp. Sci. Technol.
  28~(1) (2013) 145--153.
\newblock \href {https://doi.org/10.1016/j.ast.2012.10.010}
  {\path{doi:10.1016/j.ast.2012.10.010}}.

\bibitem{Kierkegaard_AST_2010}
A.~Kierkegaard, S.~Boij, G.~Efraimsson, A frequency domain linearized
  navier{\textendash}stokes equations approach to acoustic propagation in flow
  ducts with sharp edges, J. Acoust. Soc. Am. 127~(2) (2010) 710--719.
\newblock \href {https://doi.org/10.1121/1.3273899}
  {\path{doi:10.1121/1.3273899}}.

\bibitem{Cheng_AST_2021}
Y.~Cheng, T.~Jin, K.~Luo, Z.~Li, H.~Wang, J.~Fan, Large eddy simulations of
  spray combustion instability in an aero-engine combustor at elevated
  temperature and pressure, Aerosp. Sci. Technol. 108 (2021) 106329.
\newblock \href {https://doi.org/10.1016/j.ast.2020.106329}
  {\path{doi:10.1016/j.ast.2020.106329}}.

\bibitem{Dowling_JPP_2003}
A.~P. Dowling, S.~R. Stow, {Acoustic Analysis of Gas Turbine Combustors}, J.
  Propul. Power 19~(5) (2003) 751--764.

\bibitem{Dowling_ARFM_2005}
A.~P. Dowling, A.~S. Morgans, {Feedback control of combustion oscillations},
  Annu. Rev. Fluid Mech. 37 (2005) 151--182.

\bibitem{Han_CNF_2015}
X.~Han, A.~S. Morgans, {Simulation of the flame describing function of a
  turbulent premixed flame using an open-source LES solver}, Combust. Flame
  162~(5) (2015) 1778--1792.
\newblock \href {https://doi.org/10.1016/j.combustflame.2014.11.039}
  {\path{doi:10.1016/j.combustflame.2014.11.039}}.

\bibitem{Li_CNF_2017}
J.~Li, Y.~Xia, A.~S. Morgans, X.~Han, Numerical prediction of combustion
  instability limit cycle oscillations for a combustor with a long flame,
  Combust. Flame 185 (2017) 28--43.

\bibitem{Poinsot_PCI_2017}
T.~Poinsot, Prediction and control of combustion instabilities in real engines,
  Proc. Combust. Inst. 36~(1) (2017) 1 -- 28.
\newblock \href {https://doi.org/https://doi.org/10.1016/j.proci.2016.05.007}
  {\path{doi:https://doi.org/10.1016/j.proci.2016.05.007}}.

\bibitem{Selle_CNF_2004}
L.~Selle, G.~Lartigue, T.~Poinsot, R.~Koch, K.-U. Schildmacher, W.~Krebs,
  B.~Prade, P.~Kaufmann, D.~Veynante, {Compressible large eddy simulation of
  turbulent combustion in complex geometry on unstructured meshes}, Combust.
  Flame 137~(4) (2004) 489--505.

\bibitem{Li_OSCILOS}
J.~Li, D.~Yang, C.~Luzzato, A.~S. Morgans, {OSCILOS: the open source combustion
  instability low order simulator}, Tech. rep., Imperial College London,
  http://www.oscilos.com (2014).

\bibitem{yang2019systematic}
D.~Yang, D.~Laera, A.~S. Morgans, A systematic study of nonlinear coupling of
  thermoacoustic modes in annular combustors, J. Sound Vib. 456 (2019)
  137--161.

\bibitem{Yang_AST_2021}
L.~Yang, S.~Zhu, J.~Li, Analysis of acoustic, entropy and vorticity waves in a
  non-uniform annular combustor, Aerosp. Sci. Technol. 112 (2021) 106588.
\newblock \href {https://doi.org/10.1016/j.ast.2021.106588}
  {\path{doi:10.1016/j.ast.2021.106588}}.

\bibitem{Webster_PNAS_1919}
A.~G. Webster, Acoustical impedance and the theory of horns and of the
  phonograph, Proc. Natl. Acad. Sci. U.S.A. 5~(7) (1919) 275--282.
\newblock \href {https://doi.org/10.1073/pnas.5.7.275}
  {\path{doi:10.1073/pnas.5.7.275}}.

\bibitem{Salmon_JASA_1946}
V.~Salmon, Generalized plane wave horn theory, J. Acoust. Soc. Am. 17~(3)
  (1946) 199--211.
\newblock \href {https://doi.org/10.1121/1.1916316}
  {\path{doi:10.1121/1.1916316}}.

\bibitem{Sujith_JSV_1995}
R.~Sujith, G.~Waldherr, B.~Zinn, An exact solution for one-dimensional acoustic
  fields in ducts with an axial temperature gradient, J. Sound Vib. 184~(3)
  (1995) 389--402.
\newblock \href {https://doi.org/10.1006/jsvi.1995.0323}
  {\path{doi:10.1006/jsvi.1995.0323}}.

\bibitem{Kumar_JSV_1998}
B.~M. Kumar, R.~I. Sujith, Exact solution for one-dimensional acoustic fields
  in ducts with polynomial mean temperature profiles, J. Sound Vib. 120~(4)
  (1998) 965--969.
\newblock \href {https://doi.org/10.1115/1.2893927}
  {\path{doi:10.1115/1.2893927}}.

\bibitem{Kumar_JASA_1997}
B.~M. Kumar, R.~I. Sujith, Exact solution for one-dimensional acoustic fields
  in ducts with a quadratic mean temperature profile, J. Acoust. Soc. Am.
  101~(6) (1997) 3798--3799.
\newblock \href {https://doi.org/10.1121/1.418385}
  {\path{doi:10.1121/1.418385}}.

\bibitem{Karthik_JASA_2000}
B.~Karthik, B.~M. Kumar, R.~I. Sujith, Exact solutions to one-dimensional
  acoustic fields with temperature gradient and mean flow, J. Acoust. Soc. Am.
  108~(1) (2000) 38--43.
\newblock \href {https://doi.org/10.1121/1.429442}
  {\path{doi:10.1121/1.429442}}.

\bibitem{Veeraragavan_AIAA_2006}
A.~Veeraragavan, B.~Pesala, R.~Sujith, An integral approach to modeling sound
  propagation through a finite combustion zone, in: 44th {AIAA} Aerospace
  Sciences Meeting and Exhibit, American Institute of Aeronautics and
  Astronautics, 2006.
\newblock \href {https://doi.org/10.2514/6.2006-542}
  {\path{doi:10.2514/6.2006-542}}.

\bibitem{Li_AST_2020}
J.~Li, A.~S. Morgans, L.~Yang, The three-dimensional acoustic field in
  cylindrical and annular ducts with an axially varying mean temperature,
  Aerosp. Sci. Technol. 99 (2020) 105712.
\newblock \href {https://doi.org/10.1016/j.ast.2020.105712}
  {\path{doi:10.1016/j.ast.2020.105712}}.

\bibitem{Pagneux_JASA_1996}
V.~Pagneux, N.~Amir, J.~Kergomard, {A study of wave propagation in varying
  cross-section waveguides by modal decomposition. Part I. Theory and
  validation}, J. Acoust. Soc. Am. 100~(4) (1996) 2034--2048.
\newblock \href {https://doi.org/10.1121/1.417913}
  {\path{doi:10.1121/1.417913}}.

\bibitem{Pillai_JASA_2019}
M.~A. Pillai, D.~D. Ebenezer, E.~Deenadayalan, Transfer matrix analysis of a
  duct with gradually varying arbitrary cross-sectional area, J. Acoust. Soc.
  Am. 146~(6) (2019) 4435--4445.
\newblock \href {https://doi.org/10.1121/1.5139412}
  {\path{doi:10.1121/1.5139412}}.

\bibitem{Henrick_JASA_1983}
R.~F. Henrick, J.~R. Brannan, D.~B. Warner, G.~P. Forney, The uniform {WKB}
  modal approach to pulsed and broadband propagation, J. Acoust. Soc. Am.
  74~(5) (1983) 1464--1473.
\newblock \href {https://doi.org/10.1121/1.390148}
  {\path{doi:10.1121/1.390148}}.

\bibitem{Cummings_JSV_1977}
A.~Cummings, Ducts with axial temperature gradients: An approximate solution
  for sound transmission and generation, J. Sound Vib. 51~(1) (1977) 55--67.
\newblock \href {https://doi.org/10.1016/s0022-460x(77)80112-0}
  {\path{doi:10.1016/s0022-460x(77)80112-0}}.

\bibitem{Li_JSV_2017}
J.~Li, A.~S. Morgans, The one-dimensional acoustic field in a duct with
  arbitrary mean axial temperature gradient and mean flow, J. Sound Vib. 400
  (2017) 248--269.
\newblock \href {https://doi.org/10.1016/j.jsv.2017.03.047}
  {\path{doi:10.1016/j.jsv.2017.03.047}}.

\bibitem{Yeddula_JSV_2020}
S.~R. Yeddula, A.~S. Morgans, A semi-analytical solution for acoustic wave
  propagation in varying area ducts with mean flow, J. Sound Vib. (2020)
  115770\href {https://doi.org/10.1016/j.jsv.2020.115770}
  {\path{doi:10.1016/j.jsv.2020.115770}}.

\bibitem{Yeddula_JSV_2021}
S.~R. Yeddula, R.~Gaudron, A.~S. Morgans, Acoustic absorption and generation in
  ducts of smoothly varying area sustaining a mean flow and a mean temperature
  gradient, J. Sound Vib. 515 (2021) 116437.
\newblock \href {https://doi.org/10.1016/j.jsv.2021.116437}
  {\path{doi:10.1016/j.jsv.2021.116437}}.

\bibitem{Rani_JSV_2018}
V.~K. Rani, S.~L. Rani, {{WKB} solutions to the quasi {1-D} acoustic wave
  equation in ducts with non-uniform cross-section and inhomogeneous mean flow
  properties {\textendash} Acoustic field and combustion instability}, J. Sound
  Vib. 436 (2018) 183--219.
\newblock \href {https://doi.org/10.1016/j.jsv.2018.06.065}
  {\path{doi:10.1016/j.jsv.2018.06.065}}.

\bibitem{Basu_JSV_2022}
S.~Basu, S.~L. Rani, {Acoustic nonlinearities in a quasi 1-D duct with
  arbitrary mean properties and mean flow}, J. Sound Vib. 528 (2022) 116862.
\newblock \href {https://doi.org/10.1016/j.jsv.2022.116862}
  {\path{doi:10.1016/j.jsv.2022.116862}}.

\bibitem{Nan_AST_2021}
J.~Nan, J.~Li, Y.~Song, L.~Yang, Analytical solutions for the three-dimensional
  acoustic field in a rectangular duct with temperature gradient and mean flow,
  Aerosp. Sci. Technol. 109 (2021) 106436.
\newblock \href {https://doi.org/10.1016/j.ast.2020.106436}
  {\path{doi:10.1016/j.ast.2020.106436}}.

\bibitem{Li_JSV_2020}
J.~Li, J.~Nan, L.~Yang, Analytical solutions for the acoustic field in a thin
  annular duct with temperature gradient and mean flow, J. Sound Vib. 467
  (2020) 115043.
\newblock \href {https://doi.org/10.1016/j.jsv.2019.115043}
  {\path{doi:10.1016/j.jsv.2019.115043}}.

\bibitem{Subrahmanyam_JSV_2001}
P.~Subrahmanyam, R.~Sujith, T.~C. Lieuwen, A family of exact transient
  solutions for acoustic wave propagation in inhomogeneous, non-uniform area
  ducts, J. Sound Vib. 240~(4) (2001) 705--715.
\newblock \href {https://doi.org/10.1006/jsvi.2000.3261}
  {\path{doi:10.1006/jsvi.2000.3261}}.

\bibitem{Li_JSV_2021}
J.~Li, D.~Wang, A.~S. Morgans, L.~Yang, Analytical solutions of acoustic field
  in annular combustion chambers with non-uniform cross-sectional surface area
  and mean flow, J. Sound Vib. 506 (2021) 116175.
\newblock \href {https://doi.org/10.1016/j.jsv.2021.116175}
  {\path{doi:10.1016/j.jsv.2021.116175}}.

\end{thebibliography}
%
%

%
%
%

\end{document}